\documentclass[structabstract,a4paper]{aa}  
\usepackage{graphicx}
\usepackage{natbib}
\usepackage{array}
\usepackage{rotating}
\usepackage{txfonts}
\bibpunct{(}{)}{;}{a}{}{,}
\def\cyg{Cyg\,OB2\,\#9}
\def\l{$\lambda$}
\def\ha{H$\alpha$}
\def\hb{H$\beta$}
\def\hei{He\,{\sc i}}
\def\heii{He\,{\sc ii}}
\def\niii{N\,{\sc{iii}}}
\def\niv{N\,{\sc{iv}}}

\def\ciii{C\,{\sc{iii}}}
\def\civ{C\,{\sc{iv}}}

\def\kms{km\,s$^{-1}$}

\newcommand{\xmm}{{\sc XMM}\emph{-Newton}}
\begin{document}
   \title{The 2.35 year itch of \cyg\ }

   \subtitle{I. Optical and X-ray monitoring\thanks{Based on observations collected at OHP, with Swift, and with XMM-Newton.}}

\author{Y. Naz\'e\inst{1}\fnmsep\thanks{Research Associate FRS-FNRS}
  \and L. Mahy\inst{1}
  \and Y. Damerdji\inst{1}
  \and H.A. Kobulnicky\inst{2}
  \and J.M. Pittard\inst{3}
  \and E.R. Parkin\inst{4}
  \and O. Absil\inst{1}
  \and R. Blomme\inst{5}}

   \institute{D\'epartement AGO, Universit\'e de Li\`ege, All\'ee du 6 Ao\^ut 17, B\^at. B5C, B4000-Li\`ege, Belgium, \email{naze@astro.ulg.ac.be}
\and 
Department of Physics \& Astronomy, University of Wyoming, Laramie, WY 82071, USA
\and
School of Physics and Astronomy, The University of Leeds, Woodhouse Lane, Leeds LS2 9JT, UK
\and
Research School of Astronomy and Astrophysics, The Australian National University, Australia
\and Royal Observatory of Belgium, Ringlaan 3, B1180-Brussel, Belgium
}


 
  \abstract
   { Nonthermal radio emission in massive stars is expected to arise in wind-wind collisions occurring inside a binary system. One such case, the O-type star \cyg, was proven to be a binary only four years ago, but the orbital parameters remained uncertain. The periastron passage of 2011 was the first one to be observable under good conditions since the discovery of binarity.}
   {In this context, we have organized a large monitoring campaign to refine the orbital solution and to study the wind-wind collision.}
   {This paper presents the analysis of optical spectroscopic data, as well as of a dedicated X-ray monitoring performed with $Swift$ and \xmm.}
   {In light of our refined orbital solution, \cyg\ appears as a massive O+O binary with a long period and high eccentricity; its components (O5-5.5I for the primary and O3-4III for the secondary) have similar masses and similar luminosities. The new data also provide the first evidence that a wind-wind collision is present in the system. In the optical domain, the broad \ha\ line varies, displaying enhanced absorption and emission components at periastron. X-ray observations yield the unambiguous signature of an adiabatic collision because, as the stars approach periastron, the X-ray luminosity closely follows the $1/D$ variation expected in that case. The X-ray spectrum appears, however, slightly softer at periastron, which is probably related to winds colliding at slightly lower speeds at that time.  }
   {It is the first time that such a variation has been detected in O+O systems, and the first case where the wind-wind collision is found to remain adiabatic even at periastron passage.}

   \keywords{binaries: spectroscopic -- stars: early-type -- stars: emission-line -- stars: individual (\cyg) -- X-rays: stars}

   \maketitle
%

\section{Introduction}
Nonthermal radio emission, bright and variable hard X-ray emission, or periodic changes in the profiles of optical emission lines are the typical signatures of colliding winds in massive binary systems. Indeed, only massive stars have strong enough stellar winds for hosting these kinds of energetic phenomena. However, the reverse is not true; i.e., (many) massive binaries exist that show no signature of colliding winds. In fact, nonthermal radio emission is a rare feature (fewer than 40 cases in our Galaxy, see \citealt{deb07,ben09}), and recent X-ray surveys have shown that most binaries are not significantly overluminous at high energies \citep{naz09}. Understanding why it is so will improve our knowledge of both collision mechanisms and stellar wind properties since such emissions heavily depend on them. To this aim, it is important to collect as much information as possible, notably on the rarest cases showing conspicuous collision signatures like \cyg. 

As suggested by its name, this star belongs to the Cyg OB2 association, one of the richest OB associations of our Galaxy since it contains about one hundred O-stars \citep{kno00,neg08,wri10}. As for other members of this association, \cyg\ is strongly extinguished, which renders its study quite difficult ($E[B-V]=2.11$, \citealt{mas91}). It was nevertheless soon identified as an early-type star \citep[O5If$^+$,][]{mor55,wal73} and as a nonthermal radio emitter \citep{abb84}. After these first studies, the spectrum of \cyg\ was not studied in more detail, so that it was only proven to be a binary in recent years. A hint towards a multiple nature came from the conic shape of the nonthermal emission, a configuration revealed by the Very Large Baseline Array \citep{dou06} and only possible in wind-wind collisions. Two years later, two important findings were reported. A first, indirect evidence of binarity came from the periodic modulations of the radio emission \citep{van08}, occurring with a timescale of 2.355\,yr. Soon after that, direct evidence of binarity, i.e., clear doubling of stellar lines, was unveiled by a dedicated spectroscopic campaign \citep{naz08}. The continued monitoring of the system then led to computation of a preliminary orbital solution \citep{naz10}: \cyg\ appeared as a long-period, very eccentric ($e=0.7$) binary composed of two nearly equal-mass objects. In parallel, the analysis of several \xmm\ observations of this object revealed no strong overluminosity or peculiar hardness in the X-ray spectrum, but did show some unusual long-term variability \citep[by about 10\%, ][]{naz10}. Clearly, additional data are needed to pinpoint the properties of the system: improving the orbital solution, enlarging the orbital coverage of the X-ray data (the available ones being all taken far from periastron), and extending the knowledge of the radio emission. Only then can a full modeling be undertaken, to better understand massive stars and their winds.

Since the discovery of the binarity of \cyg, two periastron passages have occurred. The first one took place in the beginning of 2009, when the system was in conjunction with the Sun and therefore unobservable. The second one was expected to occur in June-July 2011, i.e., at the best time for observing the Cygnus constellation. We thus organized a large observing campaign of \cyg\ to observe all aspects of this event: radio monitoring of the nonthermal emission using the Expanded Very Large Array, interferometric monitoring to disentangle the astrometric orbit of the two stars using CHARA, spectroscopic monitoring in the optical to improve the  orbital solution, and X-ray monitoring to detect and understand the high-energy emission from the wind-wind collision. The results of this multiwavelength campaign will be published as a series of papers, which will be concluded by a modeling of the system. With all these observations in hand, \cyg\ will become one of the best known nonthermal radio emitters, with detailed constraints on the wind characteristics and on the collision properties. \cyg\ will thus represent a true stepping stone in our understanding of wind-wind collisions and their associated emissions. 

This first paper provides the results obtained through optical spectroscopy and X-ray monitoring. Section 2 presents the observations, Sect. 3 analyzes the spectroscopic data with the aim to improve the orbital solution, Sect. 4 discusses the variations observed in X-rays, and Sect. 5 summarizes our findings and concludes this paper.

\section{Observations}

\subsection{The optical domain}
\subsubsection{Sophie}

Our monitoring of \cyg\ at the Haute-Provence Observatory (France) began in 2007. It mainly used the high-resolution echelle spectrograph Sophie installed on the 1.93m telescope. It yields an homogeneous dataset covering a broad wavelength domain with the highest quality available. 

The Sophie spectrograph provides coverage of the 3900--6900\,\AA\ domain with a resolving power $R=\lambda/\Delta \lambda$ of 35\,000. In total, 75 echelle spectra were obtained with this instrument. The exposure time was typically 40 min, and the derived signal-to-noise ratios vary from 30 to 80. Whenever the exposures were close (i.e., they were taken within a month and no large radial-velocity [RV] variations were detected), these data were combined, providing  in the end 31 independent spectra spanning a period of four years. As for all Sophie spectra, these data were reduced by the automatic Sophie pipeline \footnote{http://www.obs-hp.fr/guide/sophie/sophie-eng.shtml}: it locates the orders, performs an optimal extraction with cosmic-ray rejection, calibrates the spectra, and finally reconnects the orders after barycentric correction and correction for the blaze function (derived from a Tungsten lamp exposure - for massive stars, it provides only a rough, first-order normalization). These data were smoothed further using a sliding window of 15\,px size (with 1\,px$=$0.01\,\AA) , and corrected by hand for any remaining cosmic ray hits. The journal of observations is provided in Table \ref{journal}.

Owing to the low signal-to-noise and the high extinction of the star, it is difficult to perform a clean and coherent normalization. To ensure homogeneity, we first performed the best normalization possible, using low-order polynomials, on the data from June 2010: these data are not only among the best quality data, but they were also taken at a time when lines are most blended, making the definition of continuum windows easier. We then used this normalized set as a reference for the other data. It enables us to remove stellar lines (through a division of the spectra by the reference), allowing better modeling with low-order polynomials of the continuum oscillations due to the poor blaze correction and poor order merging. 

In a final step, we tried to correct for some annoying nonstellar features. To this aim, we have fitted with Gaussians the diffuse interstellar bands (DIBs) in the 5412\,\AA\ and 5800\,\AA\ regions (see Fig. \ref{cordib}) on the June 2010 data (i.e., when lines are most blended). In addition, a DIB near 5812\,\AA\ was unveiled when the stellar lines were deblended - this particular DIB apparently has a similar rest wavelength to the \civ\ line. It was thus fitted on the deblended spectra and added to the set of DIBs already defined. No good fitting could be found for the DIB at 5797\,\AA, so that we kept it in the spectra. The fitted interstellar features were then removed from the data, leading to cleaner stellar lines. We have also used the table from \citet{hin00} to correct the \ha, \hei\,\l\,5876\AA, and \ciii\,\l\,5696\AA\ regions for pollution by the telluric lines. 

\onlfig{1}{
\begin{figure}[htb]
\includegraphics[width=9cm, bb=80 145 560 700, clip]{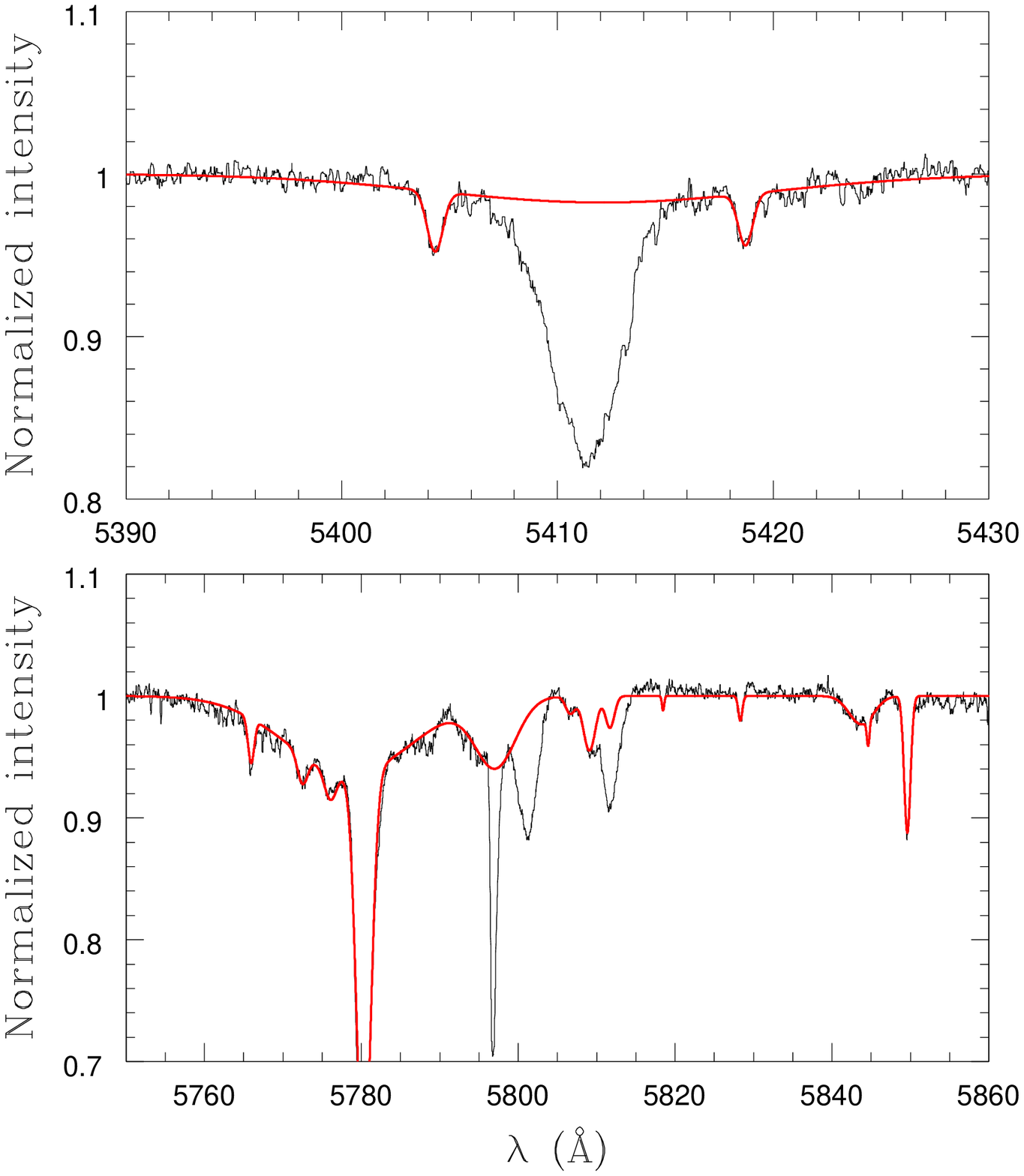}
\caption{The spectrum from June 2010 (black thin line) and the DIB correction (red thick line).}
\label{cordib}
\end{figure}
}

\subsubsection{WIRO}
Our team also monitored \cyg\ at the Wyoming Infrared Observatory (WIRO) 2.3m telescope many times between 2011 May 31 and 2011 July 25 (see Table \ref{journal}), which provides a more intense coverage of the periastron event, as well as an independent check of the stellar behavior at that time. All new datasets were reduced using standard IRAF reduction routines as outlined in \citet{kim07}. Sixteen observations were obtained using the WIRO-longslit spectrograph with the 1800~line~mm$^{-1}$ grating in first order over the spectral range 5250--6750\,\AA. Exposure times varied from 720\,s to 5400\,s depending on weather conditions; the resulting signal-to-noise was about 200 near 5876\,\AA. Copper-Argon lamp exposures were taken after each observation to wavelength-calibrate the spectra to an RMS of 0.03\,\AA. The typical resolution was 1.5\,\AA\ $FWHM$ across the chip. Spectra were Doppler corrected to the heliocentric frame and then shifted further by small amounts (typically $<6$\,\kms) so that Na\,{\sc i} interstellar absorption lines in each exposure fell at the mean Na\,{\sc i} velocity averaged over the ensemble of 16 measurements. The shift also ensures that interstellar features, e.g. the Na\,{\sc i}\,\l\,5889 line, display similar velocities in WIRO and Sophie spectra, enabling us to directly compare them. Finally, continuum normalization was performed using low-degree polynomials.

\onltab{1}{
\begin{table}[htb]
\caption{Journal of observations. }
\label{journal}
\centering
\begin{tabular}{lcc}
            \hline\hline
Date & HJD--2\,450\,000 & $\phi$ \\
\hline
Sophie\\
Sep 07  & 4348.765  &  0.379 \\
Oct 07  & 4379.063  &  0.414 \\
Dec 07  & 4463.748  &  0.512 \\
Mar 09  & 4906.675  &  0.028 \\
Apr 09  & 4936.968  &  0.064 \\
May 09  & 4953.595  &  0.083 \\
Jun 09  & 5006.497  &  0.145 \\
Jul 09  & 5022.578  &  0.164 \\
Aug 09  & 5058.433  &  0.205 \\
Sep 09  & 5079.965  &  0.230 \\
Oct 09  & 5123.370  &  0.281 \\
Nov 09  & 5142.240  &  0.303 \\
Apr 10  & 5316.621  &  0.506 \\
May 10  & 5346.054  &  0.540 \\
Jun 10  & 5372.560  &  0.571 \\
Jul 10  & 5400.557  &  0.604 \\
Aug 10  & 5433.963  &  0.643 \\
Sep 10  & 5444.420  &  0.655 \\
Oct 10  & 5482.870  &  0.700 \\
Nov 10  & 5523.761  &  0.747 \\
Dec 10  & 5557.230  &  0.786 \\
Jan 11  & 5584.273  &  0.818 \\
Mar 11  & 5626.701  &  0.867 \\
Apr 11  & 5675.632  &  0.924 \\
May 11  & 5698.603  &  0.951 \\
Jun 11a & 5731.576  &  0.989 \\
Jun 11b & 5737.249  &  0.996 \\
Jul 11a & 5745.042  &  0.005 \\
Jul 11b & 5755.550  &  0.017 \\
Jul 11c & 5769.524  &  0.034 \\
Aug 11  & 5790.465  &  0.058 \\
\hline		        
WIRO (2011)\\	        
May  31 & 5713.847  &  0.969 \\
Jun 01  & 5714.763  &  0.970 \\
Jun 03  & 5716.737  &  0.972 \\
Jun 04  & 5717.742  &  0.973 \\
Jun 05  & 5718.738  &  0.975 \\
Jun 13  & 5726.765  &  0.984 \\
Jun 14  & 5727.794  &  0.985 \\
Jun 24  & 5737.758  &  0.997 \\
Jun 25  & 5738.778  &  0.998 \\
Jun 26  & 5739.444  &  0.999 \\
Jun 27  & 5740.441  &  0.000 \\
Jun 28  & 5741.443  &  0.001 \\
Jul 15  & 5758.752  &  0.021 \\
Jul 23  & 5766.924  &  0.031 \\
Jul 24  & 5767.438  &  0.031 \\
Jul 25  & 5768.452  &  0.032 \\
\hline		        
\xmm \\		        
Oct 04  & 3308.580  &  0.167 \\
Nov 04a & 3318.558  &  0.178 \\
Nov 04b & 3328.544  &  0.190 \\
Nov 04c & 3338.506  &  0.202 \\
Apr 07  & 4220.355  &  0.229 \\
May 07  & 4224.170  &  0.233 \\
Jun 11  & 5738.255  &  0.997 \\
\hline		        
{\it Swift} \\	        
Jan 11  & 5571.618  &  0.803 \\
Apr 11  & 5655.837  &  0.901 \\
May 11  & 5700.082  &  0.953 \\
Jul 11  & 5743.839  &  0.004 \\
Oct 11  & 5842.169  &  0.118 \\
\hline
\end{tabular}
\tablefoot{Datasets are identified by the month of observation followed by the last two digits of the year, except for the WIRO data where the numbers correspond to the dates.}
\end{table}
}

\subsection{X-rays}
\subsubsection{{\it Swift}}
The {\it Swift} facility observed \cyg\ during 5\,ks at five carefully selected dates in 2011 (PI Naz\'e, see Table \ref{journal}), to follow the star during ten months around the date of periastron. The data were then reduced following the recommendations of the UK {\it Swift} center\footnote{http://www.swift.ac.uk/analysis/xrt/}. First, the pipeline (FTOOLS  task {\it xrtpipeline}) was applied. Events were then extracted in a circular region of 47" radius for  the source and in a nearby circle of 119" radius for the background. The chosen region is as  devoid of X-ray sources as possible (see \citealt{rau11}). The adequate response matrix file (RMF, swxpc0to12s6\_20010101v013.rmf) is provided by the {\it Swift} team, but the ancillary response file (ARF) was specifically calculated for \cyg\ by the task {\it xrtmkarf} with the inclusion of an exposure map so that bad columns can be taken into account. Count rates measured in these observations vary from 0.04 to 0.1\,cts\,s$^{-1}$, so there is no pile-up. 

\begin{figure*}[htb]
\centering
\includegraphics[width=16cm]{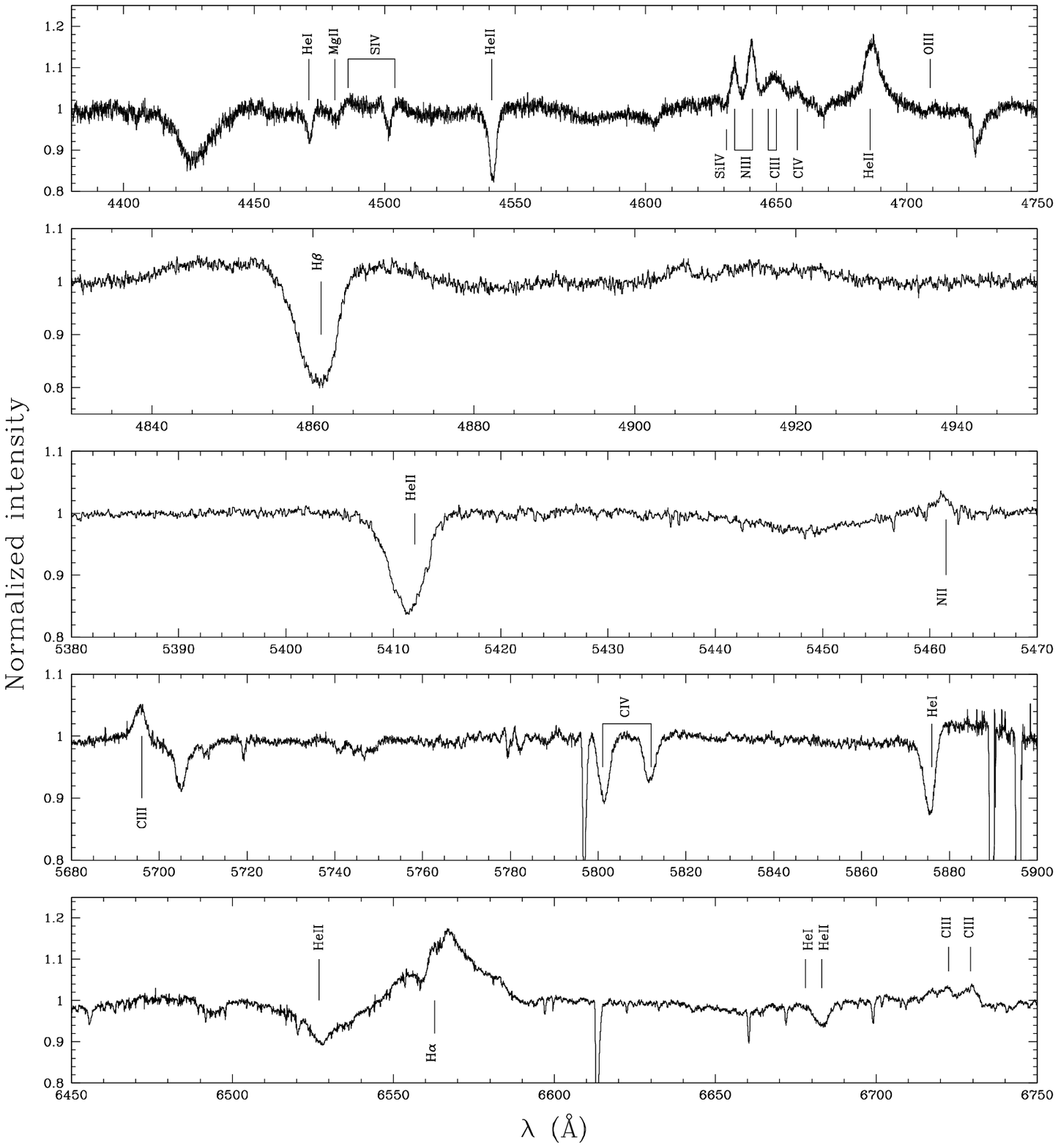}
\caption{Spectrum observed in June 2010 (after DIB and telluric correction). The labels indicate the stellar lines (other features are interstellar). The \heii\,\l\,6560\AA\ line next to \ha\ is strongly blended with a DIB. }
\label{spec}
\end{figure*}

\subsubsection{\xmm}
\xmm\ had observed the Cygnus OB2 region six times in previous years, mostly for studying the X-ray emission of Cyg\,OB2\,\#8A (see Table \ref{journal}). Since \cyg\ was in the field-of-view, we analyzed its X-ray emission and found some unexplained variations \citep{naz10}. A full monitoring of \cyg\  with \xmm\ in 2011 was not possible, but an observation was scheduled near the predicted time of the periastron passage (25\,ks, ObsID=0677980601, PI Naz\'e, see also Table \ref{journal}). 

This new dataset was reduced in the standard way using SAS v. 10.0.0. After the pipeline processing (SAS tasks {\it emproc, epproc}), we filtered the data to only keep the best events, as recommended  by the SAS team\footnote{See threads available on \\ http://xmm.esac.esa.int/sas/current/documentation/threads. We kept single and double events for MOS ($PATTERN$=0--12 and {\it XMMEA\_EM} filter) and single events for pn (null  $PATTERN$ and $FLAG$=0--4).}. A flare in the X-ray background affects the middle of the observation. It was eliminated, so that the effective exposure times are reduced to 24\,ks for MOS and 15\,ks for pn.

Source events were extracted in a circular region of 35" radius around the target, and the background region was defined as a nearby circle of the same size. An annulus could not be used, as numerous X-ray sources can be found in the region (see \citealt{rau11}). The count rate measured for this new observation is 0.6\,cts\,s$^{-1}$ for MOS and 1.8\,cts\,s$^{-1}$ for pn, which is five times higher than in previous observations. Despite this brightening, the source is not affected by pile-up, as checked by a run of the task {\it epatplot}.

\section{\cyg, as seen in the optical}

\subsection{The spectra}
The high extinction of \cyg\ has two direct consequences on its optical spectrum. First, the blue part of the spectrum ($<4400$\,\AA) is very noisy, hence not really usable. Second, most lines recorded in the spectra actually are interstellar features! Only a few stellar lines can be distinguished and used for further study (see Fig. \ref{spec}): Balmer lines (\ha, \hb), \hei\,\l\,4471, 5876\AA, \heii\,\l\,4542, 4686, 5412, 6683\AA, \niii\,\l\,4634, 4641\AA\ (in emission), \ciii\,\l\,4650, 5696, 6723, 6729\AA\ (in emission), \civ\,\l\,5801, 5812\AA. Some of these stellar lines are, however, contaminated by interstellar features, such as C\,{\sc iv}\,\l\,5812\AA.

\begin{figure}
\includegraphics[width=9.5cm]{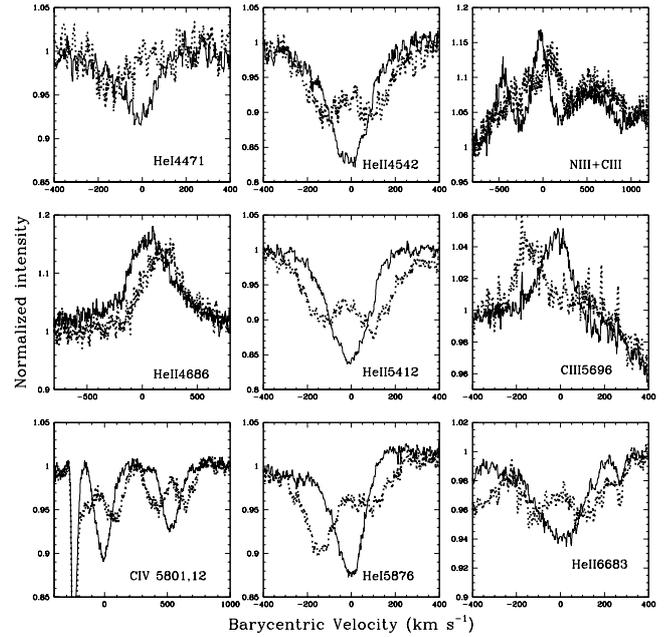}
\caption{Comparison of the spectral appearance of \cyg\ in June 2010 ($\phi$=0.571, solid line) and in the beginning of July 2011 ($\phi$=0.005, thick dotted line). The spectra have been corrected for some DIBs (see text) and the telluric lines (the latter correction being imperfect around 5696\,\AA). }
\label{compa}
\end{figure}

\onlfig{4}{
\begin{figure*}
\includegraphics[width=15cm]{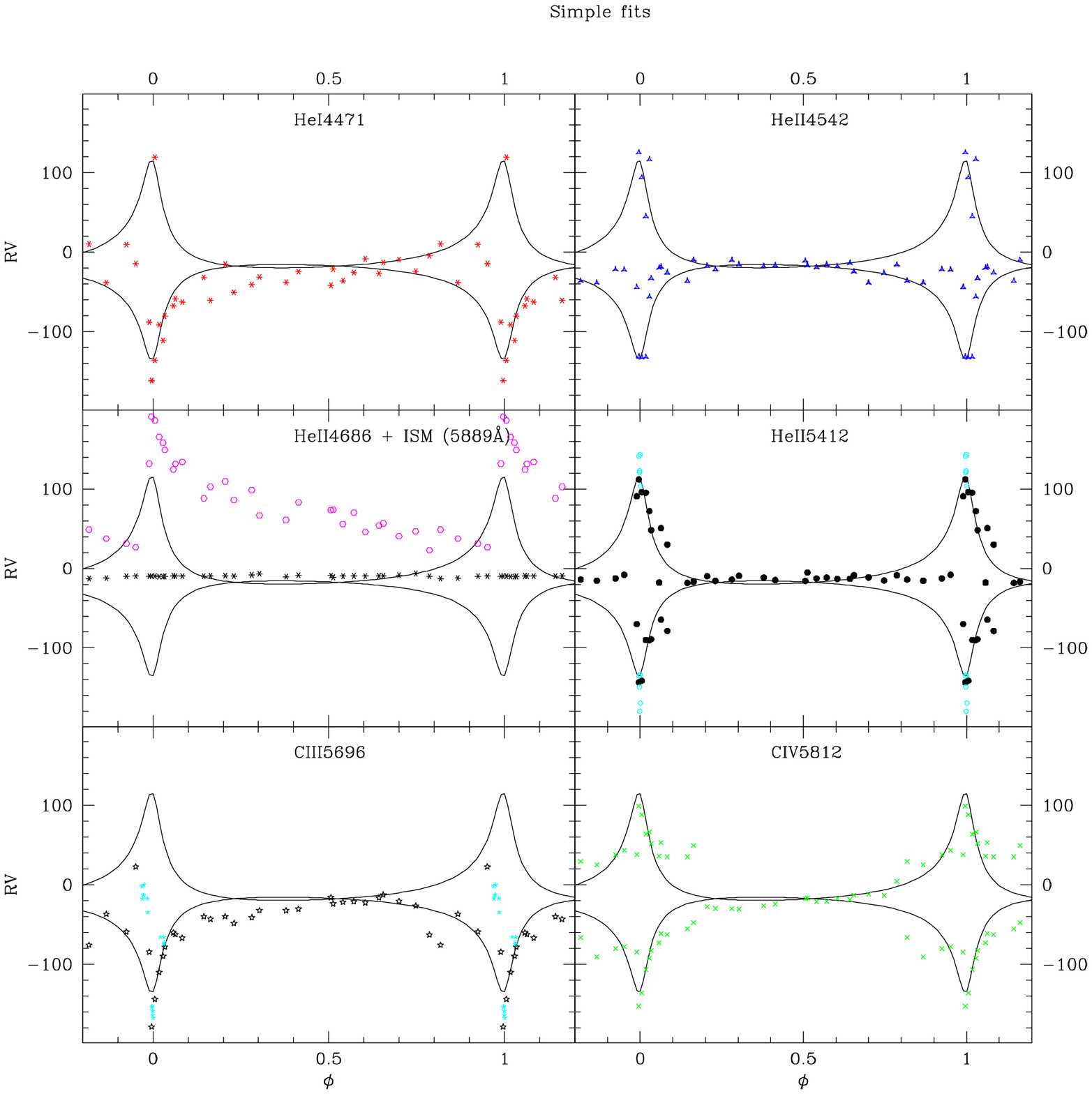}
\includegraphics[width=9cm]{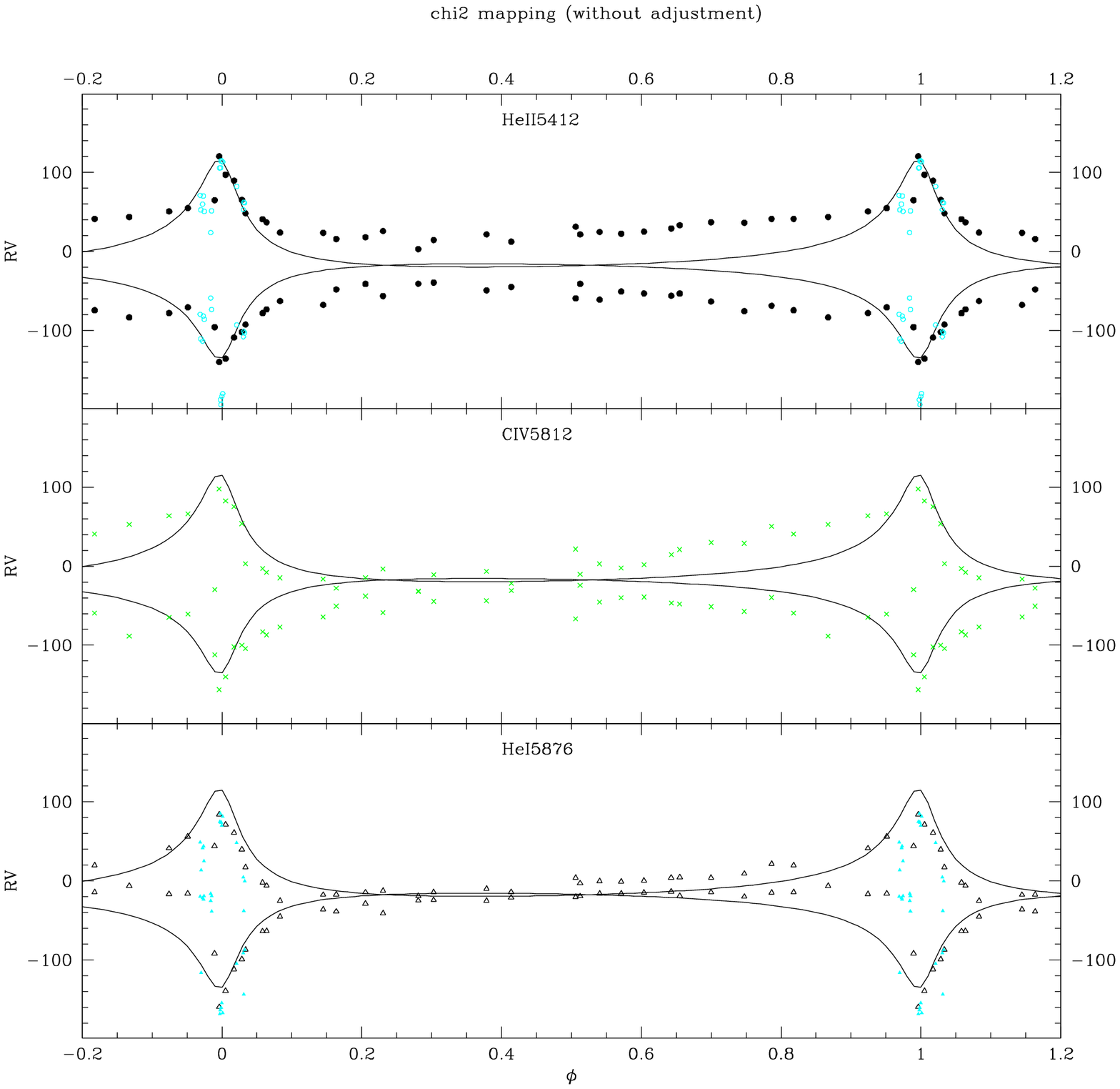}
\includegraphics[width=9cm]{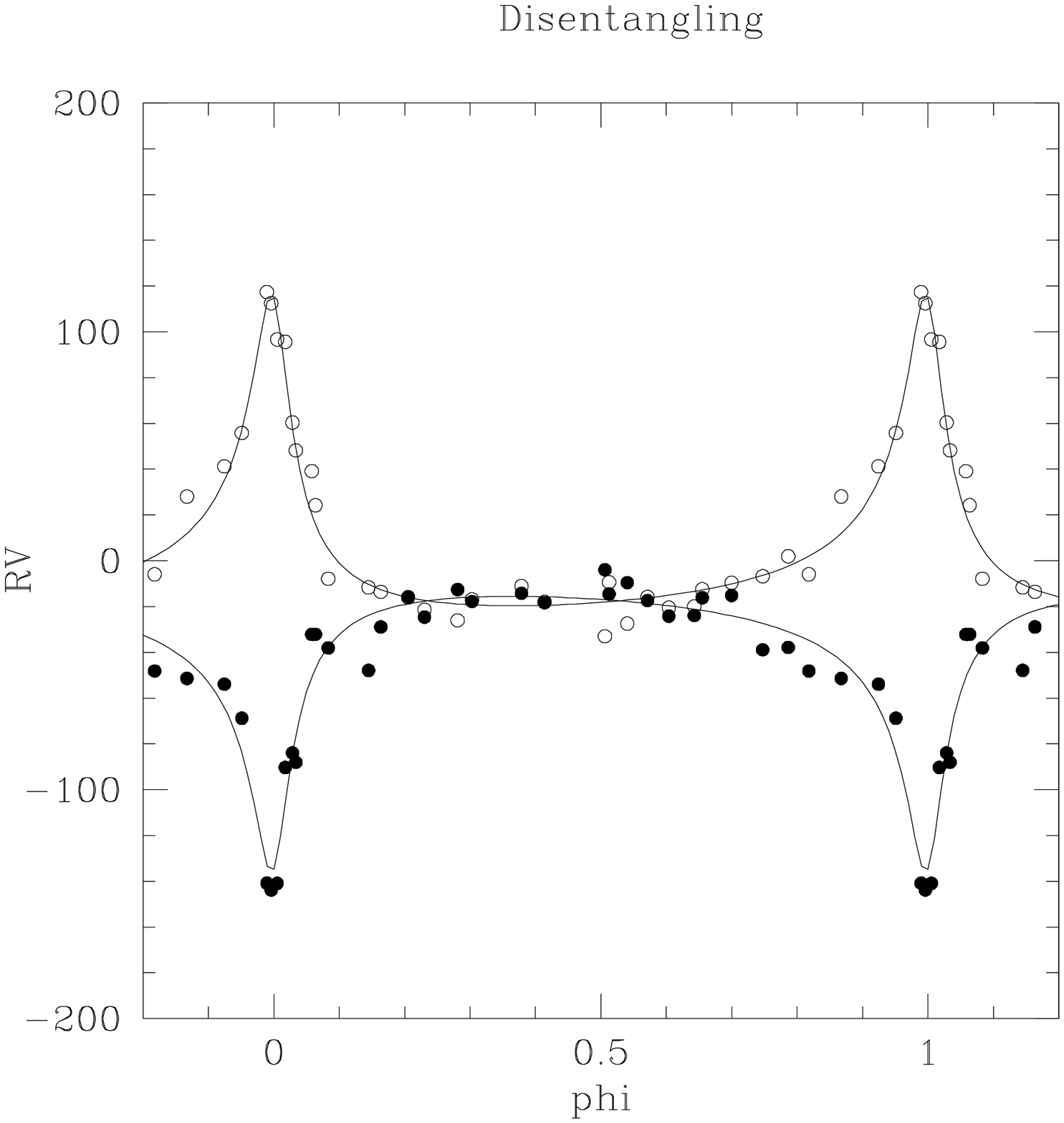}
\caption{RVs determined by simple fits (top) and by finding the best shift of fixed Gaussians (bottom left). Small cyan symbols correspond to WIRO data, other symbols are the same in all panels, with one colored symbol for each line. The RV measured for a narrow interstellar feature is superimposed on the \heii\,\l\,4686\AA\ subpanel using asterisks, to show the quality of the wavelength calibration. The bottom right panel shows the RVs determined by disentangling. The best-fit orbital solution (Table \ref{solorb}) found from disentangling is superimposed on each panel.}
\label{rvcurve}
\end{figure*}
}

When one compares the spectra taken at different phases (see Fig. \ref{compa}), it is obvious that the \ciii\,\l\,5696\AA\ emission line follows the motion of one component of the system. The \heii\,\l\,4686\AA\ emission line instead displays the opposite motion, that of the other component, though with a shift of about 90\,\kms. This shift implies that the line most probably does not arise from the photosphere, but somewhere close to it, in the wind. The exact position may be influenced by the presence of a wind-wind collision in the system (see below). These two lines are the sole ones ``associated'' to a single component of the system. Other lines show two components in the 0.9--1.1 phase interval (Fig. \ref{compa}). Another problem is obvious when one looks at the spectra where both components are clearly separated: the lines of both objects have similar strengths in \heii, \niii, and \civ, which does not help the derivation of the RVs and their attribution to one of the components. 

\onlfig{5}{
\begin{figure*}[htb]
\centering
\includegraphics[width=8cm]{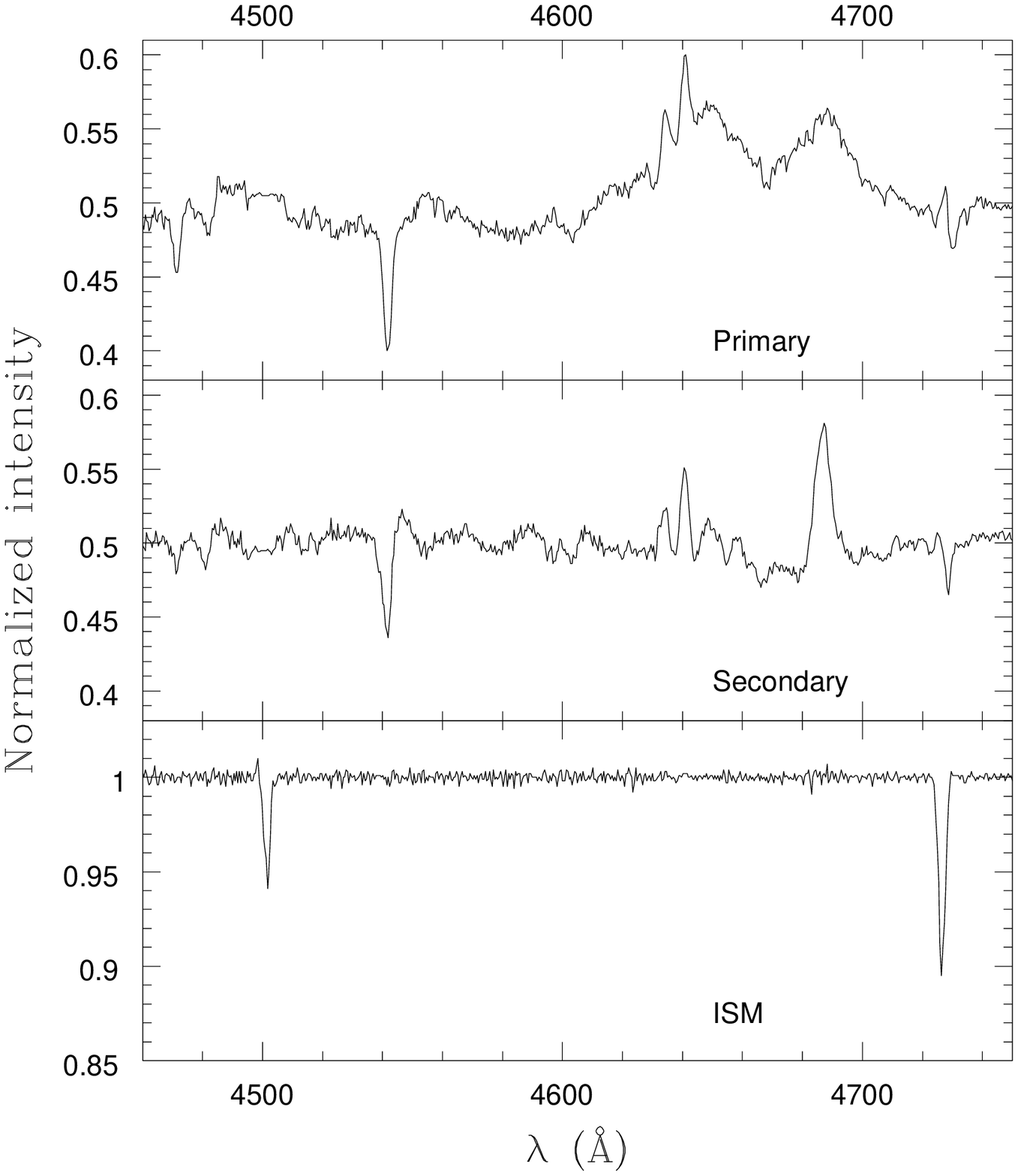}
\includegraphics[width=8cm]{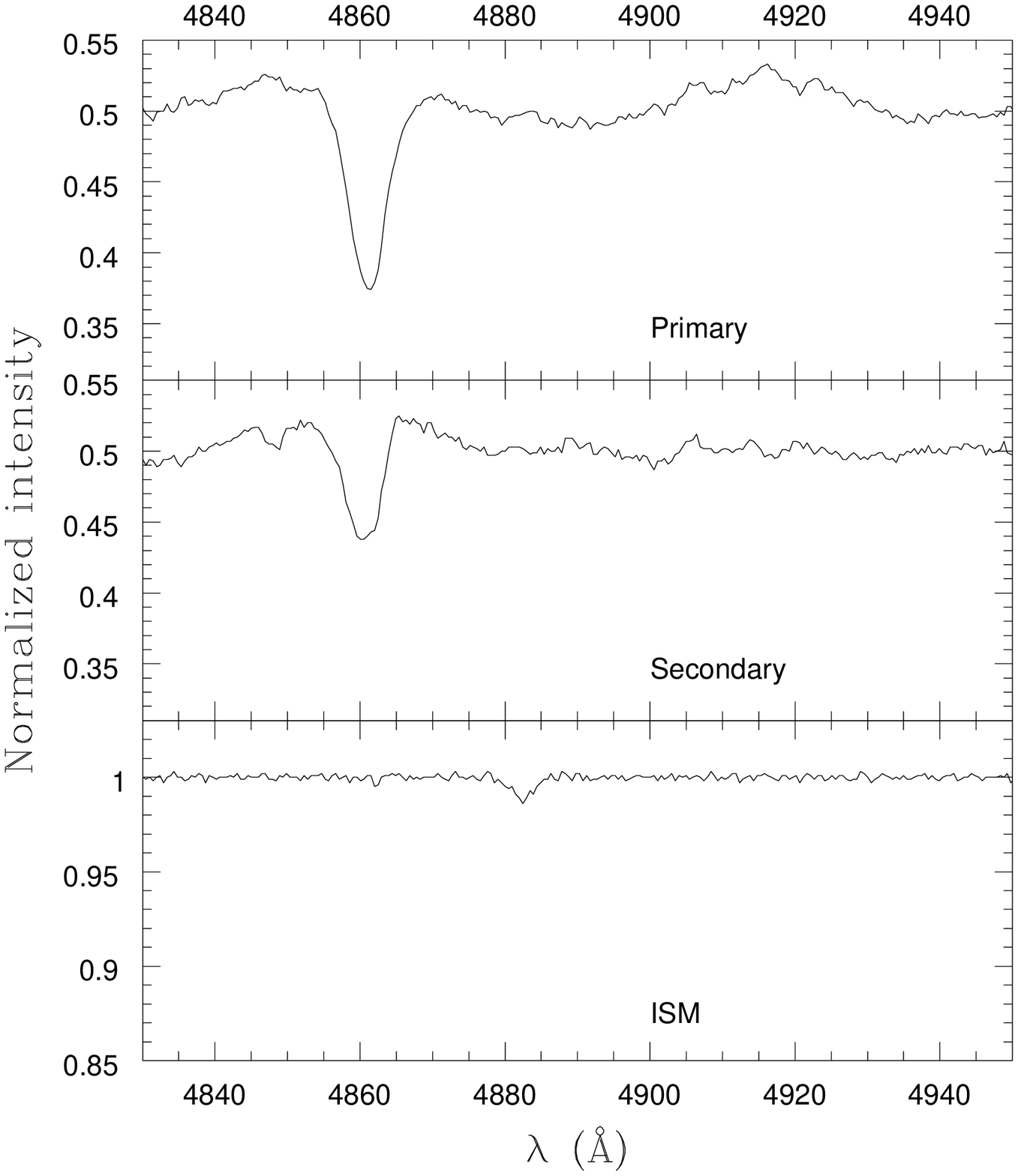}
\includegraphics[width=8cm]{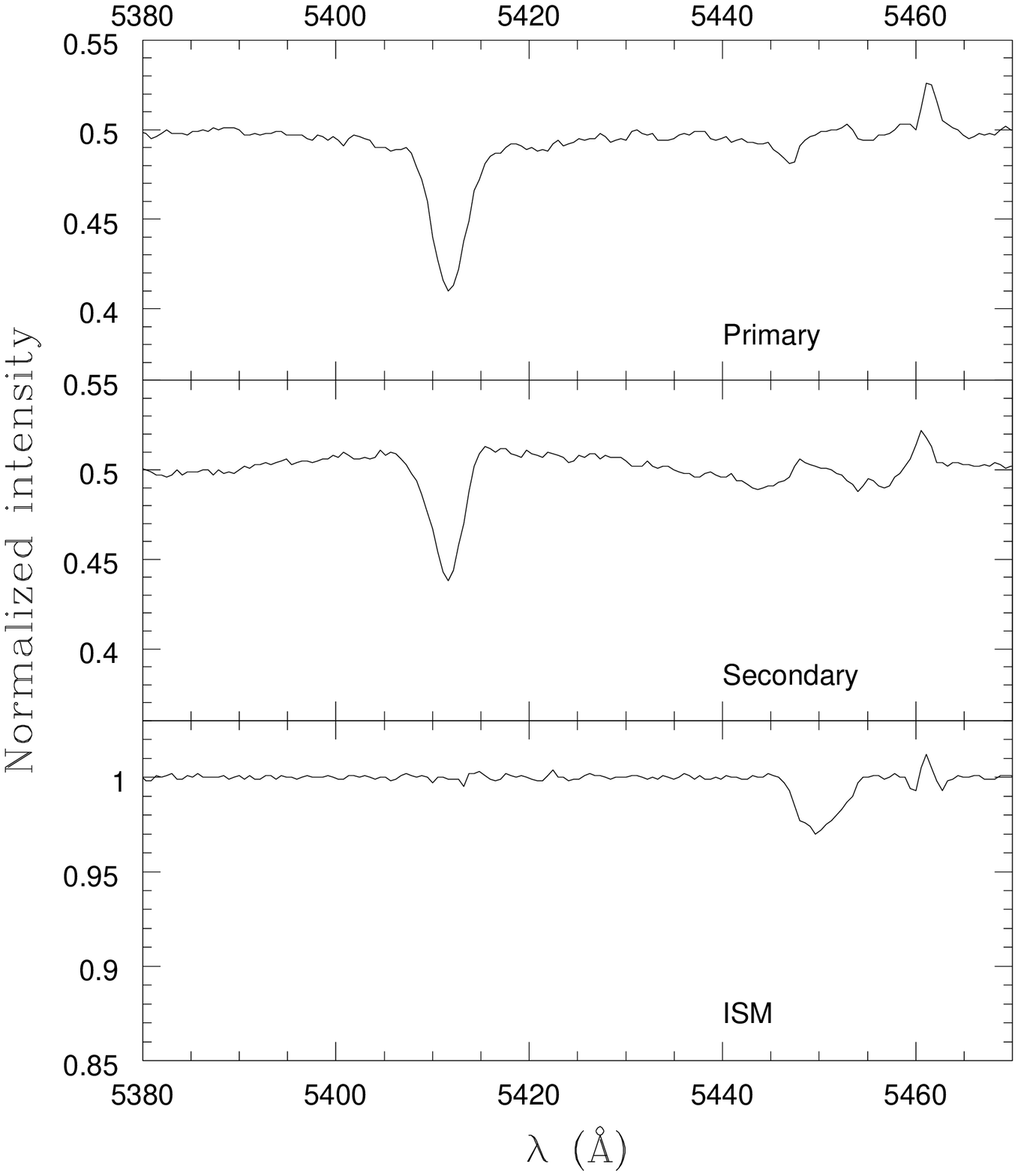}
\includegraphics[width=8cm]{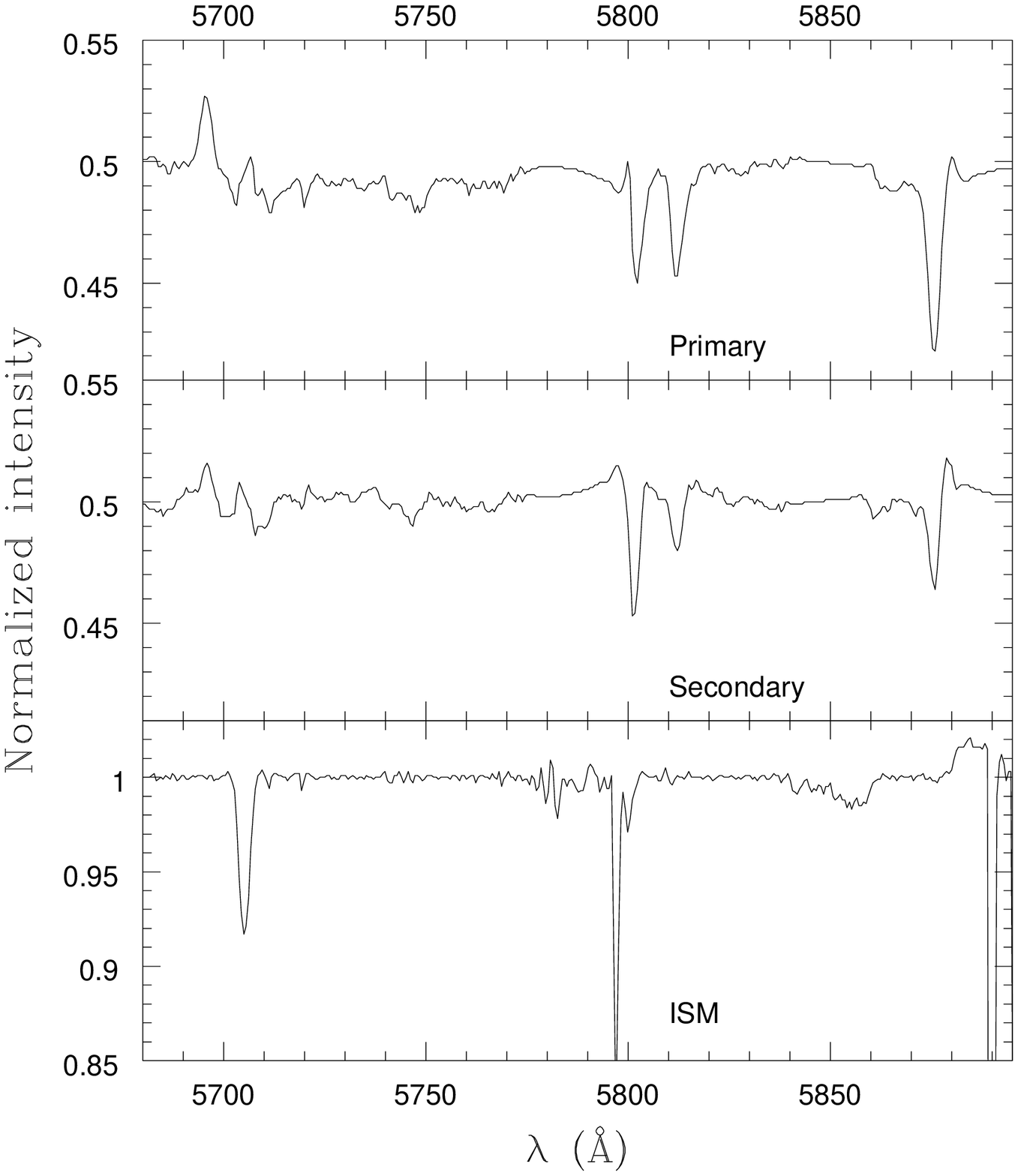}
\caption{Disentangling results. Each panel shows, from top to bottom, the spectra of the primary star, the secondary star, and the interstellar features (with continuum shifted to +1) in four different spectral domains. The normalization is imperfect, as is often the case for disentangling results when the orbital RV shifts are modest.}
\label{disent}
\end{figure*}
}
\begin{figure}
\includegraphics[width=9cm,bb=20 175 545 700, clip]{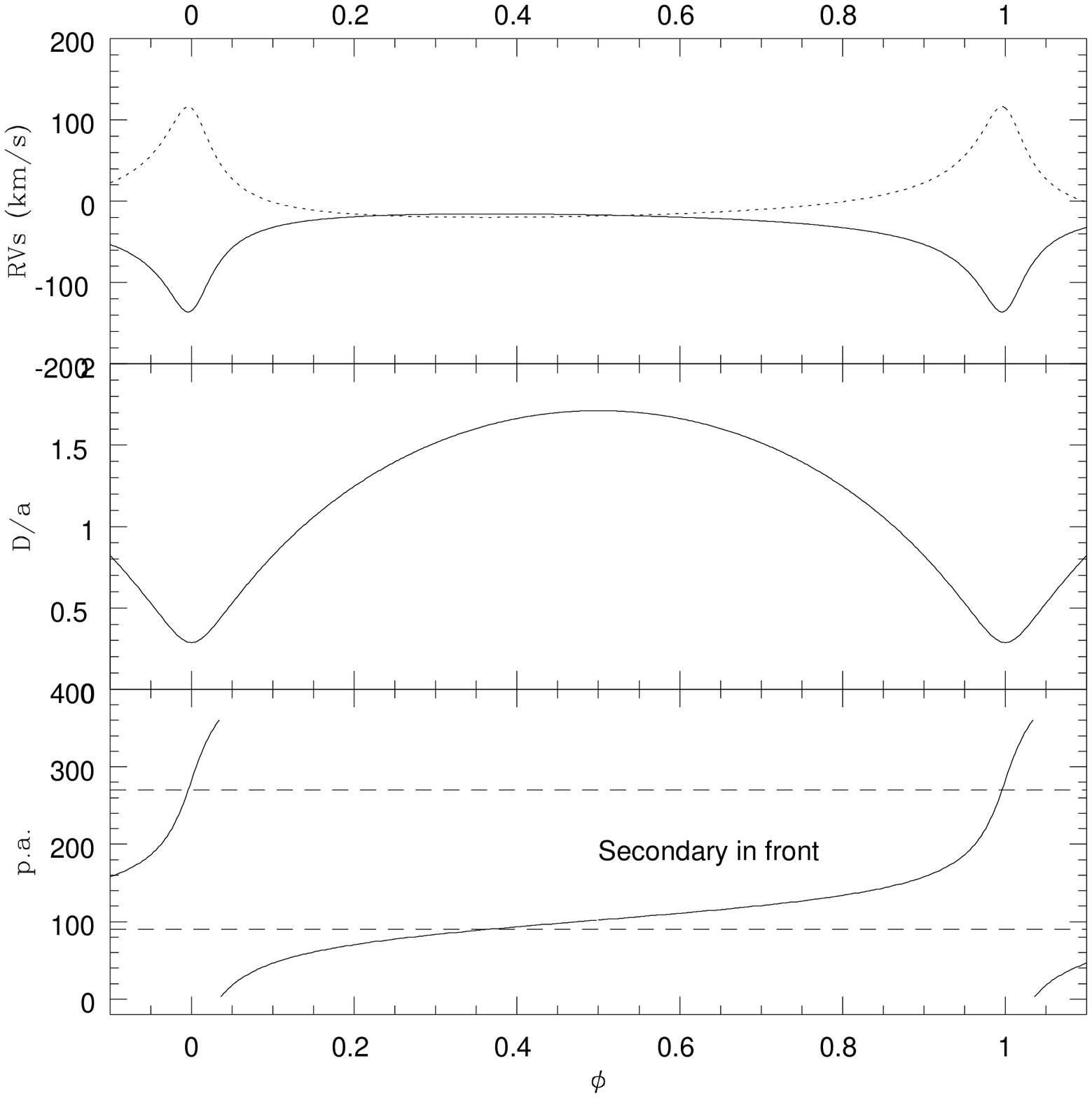}
\includegraphics[width=9cm, bb=40 170 520 640, clip]{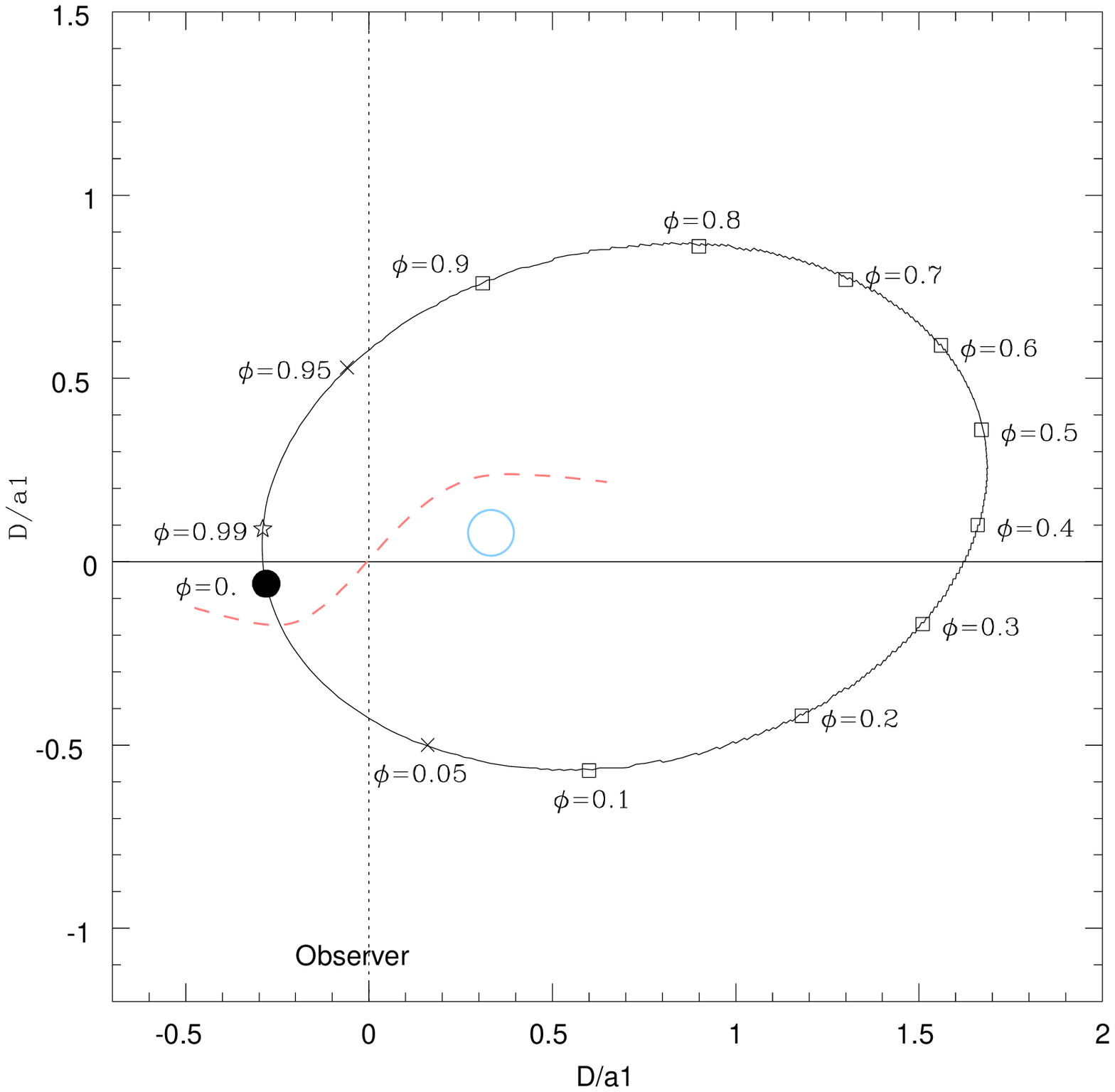}
\caption{{\it Top:} Variation with phase of the RVs (solid line=primary, dashed line=secondary), separation, and position angle (defined to be zero when the primary is in front and along the line-of-sight). {\it Bottom:} The orbit of the primary star around the center of mass (see Table \ref{solorb} for physical parameters). The position of periastron is shown with a big dot, of phases 0.1 to 0.9 with open squares, of phases 0.05 and 0.95 with crosses, and of phase 0.99 with a star. The line of nodes corresponds to the x-axis and the line-of-sight to the y-axis, with an Earthly observer towards the bottom. At periastron, a (arbitrary) sketch of the collision zone is shown as a pink dashed line, with the position of the secondary star shown as an open blue circle.}
\label{sketch}
\end{figure}

\subsection{RV measurements and orbital solution}
As already explained in \citet{naz10}, measuring the RVs in \cyg\ is difficult because the broad stellar lines are faint, sometimes even contaminated by interstellar lines, and also blended for most of the (highly eccentric) orbit. In a first step, we measured the RVs by fitting one (or two, if the line displayed two components or appeared asymmetrical, which happens within $\Delta \phi=0.1$ of periastron) Gaussian(s) to the stellar lines. Results of these simple fits are shown at the top of Fig. \ref{rvcurve}. It is obvious from that figure that deblending the two components is not easy, since they can only be measured at maximum separation. The shift of \heii\,\l\,4686\AA\ with respect to the true stellar motion is also obvious, as is the attribution to the primary star of \ciii\,\l\,5696\AA\footnote{The WIRO measures of this line around $\phi$=0.9 are very uncertain due to the unexplained presence of a faint, broad second component (normalization problem?).} and the main component of \hei\ lines. WIRO data agree well with the Sophie observations, when taking their larger uncertainty (about 20--30\,\kms) due to a lower resolution into account.

To improve over these simple fits, we fixed the component's shapes to the Gaussians fitted in the most separated cases, and we cross-correlated them with the blended lines until the best $\chi^2$ was found, as we did in \citet{naz10}. The new RV curves are better, as can be seen in Fig. \ref{rvcurve}. However, a trend must be noted: the $\chi^2$ mapping has a tendency to separate the components more at phases of maximum blending. This may be due to slight changes in line shapes (due to noise and imperfect normalization), but allowing the line shapes to vary slightly produces erratic results of lower quality.

Finally, we used these preliminary RVs as input for a disentangling program based on the method of \citet{gon06}. In this program, the spectra of each component are iteratively determined, first on the basis of the approximate input RVs, then by using improved RVs found from cross-correlating the components' spectra with the observed spectra (for details see \citealt{mah10}). In our case, the disentangling tries to fit the spectra of three components: the fixed interstellar features, and the primary and secondary stars (whose RVs vary from one observation to the next). The continuum was arbitrarily fixed at 0, 0.5, and 0.5 for these three components (i.e., the true light ratios are not taken into account in this method). The cross-correlation was performed using masks, containing the \hei\,\l\,4471\AA, \heii\,\l\,4542\AA, \heii\,\l\,5412\AA, and \hei\,\l\,5876\AA\ lines for the primary and the \heii\,\l\,4542\AA, \heii\,\l\,5412\AA, and \hei\,\l\,5876\AA\ lines for the secondary. The spectra where both components are clearly visible (i.e., maximum separation) received a weight 50 times greater than in the blended cases. This process yields the RV curves seen in the last panel of Fig. \ref{rvcurve} (for values, see Table 2), and the disentangled spectra are shown in Fig. \ref{disent}. 

\begin{figure}
\includegraphics[width=9cm]{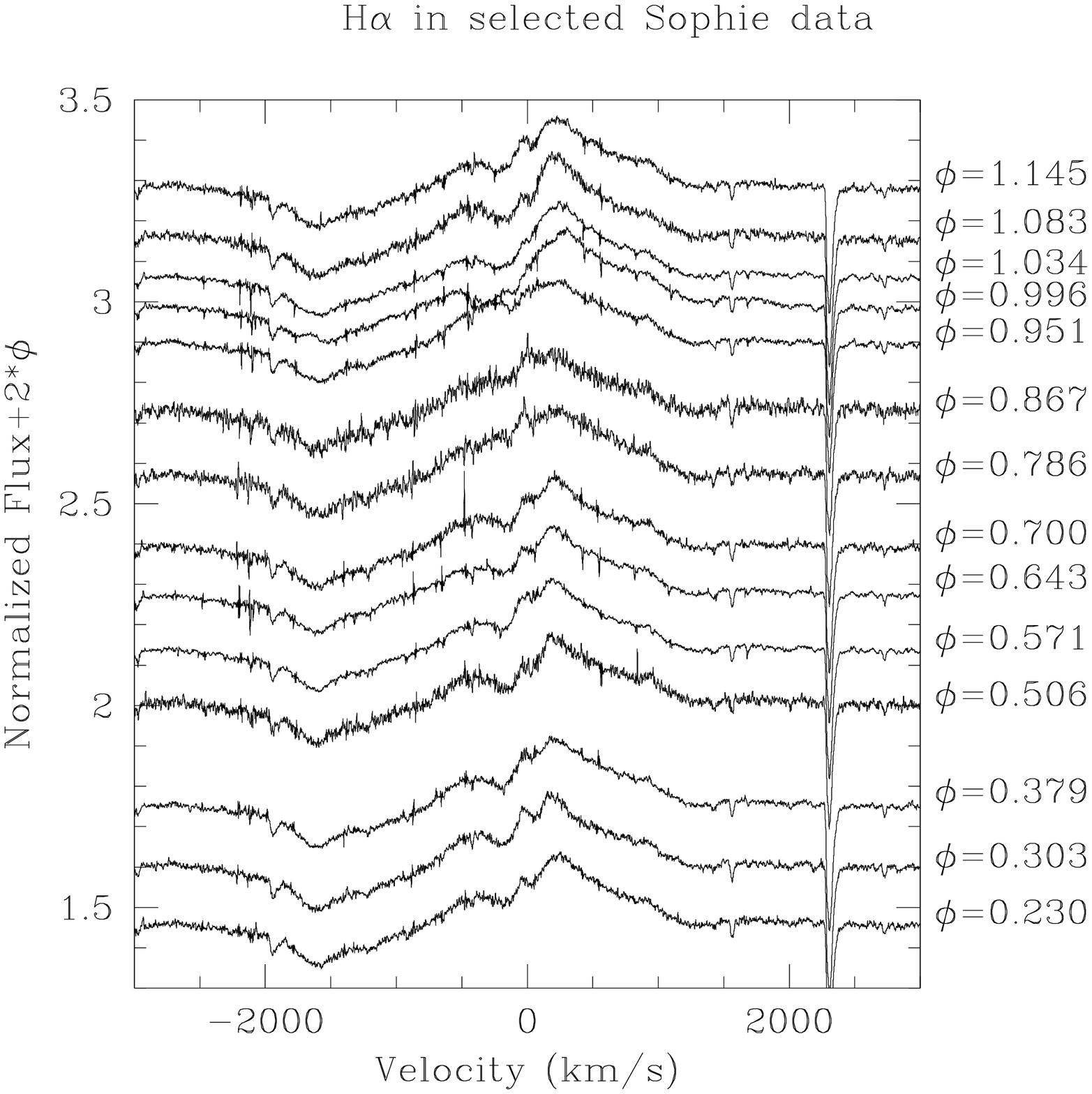}
\includegraphics[width=9cm]{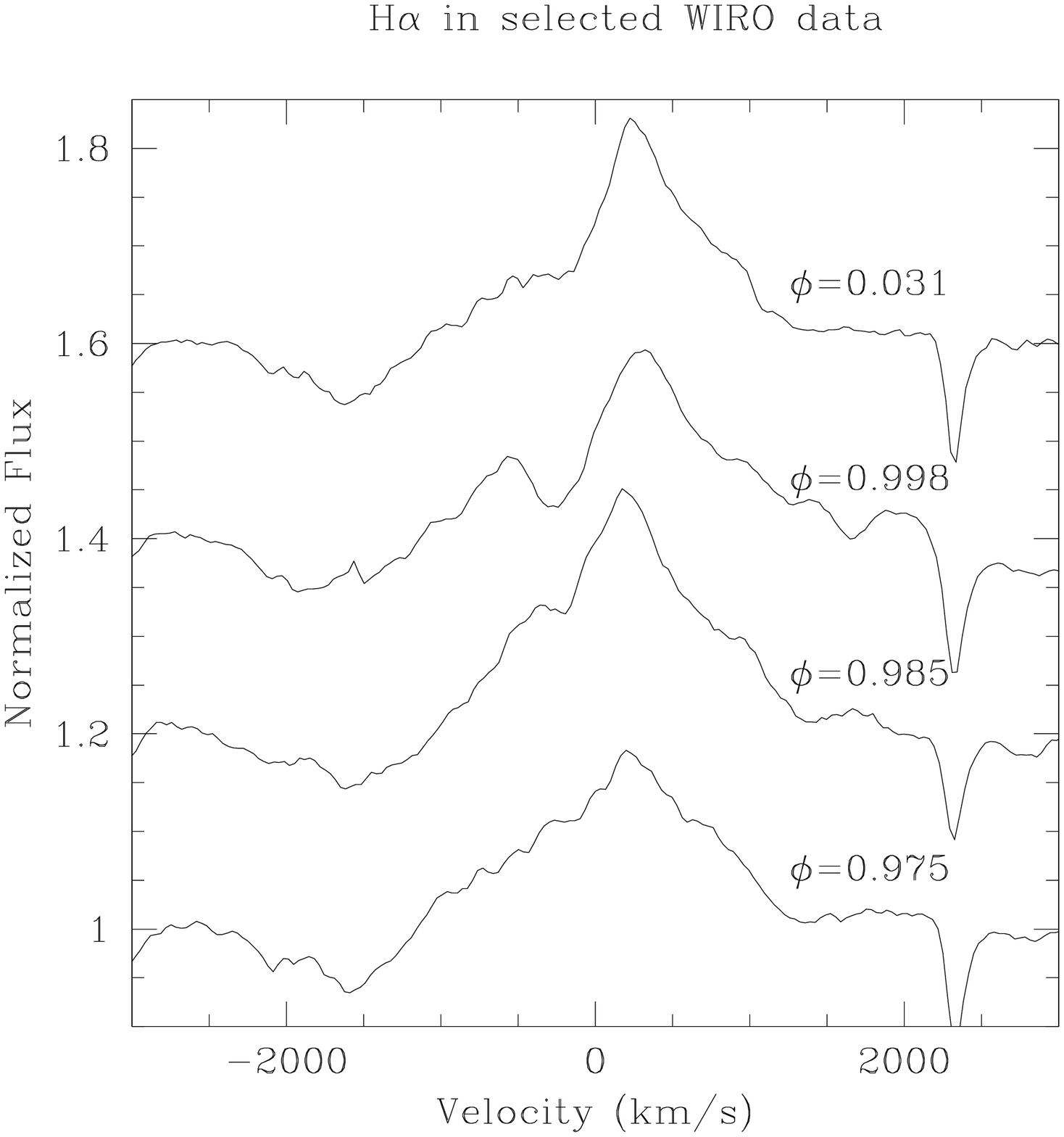}
\caption{Evolution of the \ha\ line in the Sophie (top) and WIRO (bottom) datasets. Phases are quoted near each spectrum and, in the top panel, the distance between spectra is proportional to the phase difference. Note the two absorptions that disappear before periastron, the line profile then being almost triangular, and the increased absorption and emission components at periastron.}
\label{ha}
\end{figure}

\onltab{2}{
\begin{table}[htb]
\caption{Radial velocities found from disentangling for the Sophie data.}
\label{tablervt}
\centering
\begin{tabular}{lccc}
            \hline\hline
Date & HJD--2\,450\,000 & $RV_1$ &  $RV_2$\\
 & & \kms & \kms \\
\hline
Sep 07  & 4348.765  &  $-$14.2  &    $-$11.1  \\
Oct 07  & 4379.063  &  $-$18.3  &    $-$17.9  \\
Dec 07  & 4463.748  &  $-$14.5  &     $-$9.3  \\
Mar 09  & 4906.675  &  $-$83.9  &     60.4    \\
Apr 09  & 4936.968  &  $-$32.1  &     24.3    \\
May 09  & 4953.595  &  $-$38.1  &     $-$7.9  \\
Jun 09  & 5006.497  &  $-$47.9  &    $-$11.6  \\
Jul 09  & 5022.578  &  $-$28.9  &    $-$13.5  \\
Aug 09  & 5058.433  &  $-$15.7  &    $-$16.0  \\
Sep 09  & 5079.965  &  $-$24.6  &    $-$21.4  \\
Oct 09  & 5123.370  &  $-$12.6  &    $-$26.0  \\
Nov 09  & 5142.240  &  $-$17.8  &    $-$16.9  \\
Apr 10  & 5316.621  &   $-$4.0  &    $-$33.0  \\
May 10  & 5346.054  &   $-$9.5  &    $-$27.4  \\
Jun 10  & 5372.560  &  $-$17.4  &    $-$15.8  \\
Jul 10  & 5400.557  &  $-$24.2  &    $-$20.5  \\
Aug 10  & 5433.963  &  $-$23.9  &    $-$19.9  \\
Sep 10  & 5444.420  &  $-$16.2  &    $-$12.5  \\
Oct 10  & 5482.870  &  $-$15.1  &     $-$9.6  \\
Nov 10  & 5523.761  &  $-$38.9  &     $-$6.8  \\
Dec 10  & 5557.230  &  $-$37.8  &      2.0    \\
Jan 11  & 5584.273  &  $-$48.2  &     $-$5.9  \\
Mar 11  & 5626.701  &  $-$51.4  &     28.1    \\
Apr 11  & 5675.632  &  $-$53.9  &     41.2    \\
May 11  & 5698.603  &  $-$68.7  &     55.8    \\
Jun 11a & 5731.576  & $-$140.9  &    117.4    \\
Jun 11b & 5737.249  & $-$143.8  &    112.5    \\
Jul 11a & 5745.042  & $-$140.9  &     96.7    \\
Jul 11b & 5755.550  &  $-$90.3  &     95.6    \\
Jul 11c & 5769.524  &  $-$88.0  &     48.2    \\
Aug 11  & 5790.465  &  $-$32.2  &     39.1    \\
\hline
\end{tabular}
\end{table}
\clearpage
}

The RVs derived from the disentangling can be used to find the orbital solution. To this aim, we used the Li\`ege Orbital Solution Package (LOSP)\footnote{The LOSP package is maintained by H. Sana. It is available, with a preprint describing it, from http://staff.science.uva.nl/$\sim$hsana/losp.html.}, which is based on the SB1 method of \citet{wol67} generalized to SB2 cases as described, e.g., in \citet{rau00}. The period was fixed to 858.4d, as found from an extensive radio dataset covering three decades (Blomme et al., in preparation)\footnote{The time interval covered by the spectroscopic data, about 1500\,d, is too short for a precise period determination.}, and the derived parameters are shown in Table \ref{solorb} and Fig.\ref{sketch} . It agrees well with the preliminary solutions from \citet{naz10}.

\begin{table}
\caption{Orbital solution for \cyg. $T_0$ (hence $\phi=0$) corresponds to periastron passage, in $HJD-2\,450\,000$.}
\label{solorb}
\centering
\begin{tabular}{lc}
            \hline\hline
Parameter & Value \\
            \hline
$P$(d) - from radio       & $858.4\pm1.5$   \\ 
$T_0$                     & $4023.821\pm2.51$  \\  
$e$                       & $0.713\pm0.016$ \\  
$\omega_1$($^{\circ}$)      & $192.1\pm2.9$  \\
$M_1/M_2$                 & $1.13\pm0.08$    \\ 
$\gamma_1$(\kms)          & $-33.8\pm2.7$  \\ 
$\gamma_2$(\kms)          & $0.9\pm2.8$     \\
$K_1$(\kms)               & $60.3\pm3.0$     \\
$K_2$(\kms)               & $68.0\pm3.4$     \\
$a_1\sin i$(R$_{\odot}$)  & $716.9\pm39.5$   \\
$a_2\sin i$(R$_{\odot}$)  & $808.4\pm44.5$   \\ 
$M_1\sin^3 i$(M$_{\odot}$)& $34.3\pm4.6$     \\ 
$M_2\sin^3 i$(M$_{\odot}$)& $30.4\pm4.0$     \\ 
RMS(\kms)                 & 9.6             \\  	
\hline
\end{tabular}
\end{table}

\subsection{Spectral features}

While photospheric \heii\ absorption lines display similar strength in both components (Fig. \ref{compa}), the \hei\ lines appear much stronger in the primary component (i.e., the one that appears blueshifted at periastron). In fact, \hei\,\l\,4471\AA\ is so weak in the other component that it is difficult to distinguish it. Comparing the relative strength of \heii\,\l\,4542\AA\ and \hei\,\l\,4471\AA\ \citep{con73}, we found that they correspond to an O5--5.5 type and O3--4 type for the primary and secondary, respectively (\citealt{wal10} reported the blended spectrum as O4.5Ifc). That one component is hotter than the other one may explain why \ciii\,\l\,5696 is best seen in the cooler component while \civ\ lines have similar strengths in both stars. Since the \heii\ lines display similar intensities, it also suggests that the two stars should have different luminosity classes, the coldest object being also bigger. This conclusion is supported by the fact that identical luminosity classes would also be incompatible with the coldest star being the most massive component (see previous section).

To investigate this issue further, we turned to other spectral features. Both components display \niii\ and the neighboring \ciii\ lines in emission. The latter lines are not variable and are much stronger in the primary, which agrees well with the finding that these features reach their maximum strength in O5 stars \citep{wal10}. In our disentangled spectra, the \heii\,\l\,4686\AA\ line appears in emission in both components, but the width of the feature in the primary spectrum suggests that this line may be contaminated. Indeed, the disentangling program is not fit to study such features since it considers the line shapes to be constant throughout the orbit and to arise solely in the stellar atmospheres. The shift observed for the \heii\,\l\,4686\AA\ line (see above) clearly indicates that it is not the case: even if the stars might truly display some \heii\,\l\,4686\AA\ in emission, most of it should be formed in the wind, close to the secondary star and with a potential influence of the wind-wind collision. With \niii\ + \ciii\ lines (and possibly \heii\,\l\,4686\AA) in strong emission, the primary spectrum thus appears quite compatible with a supergiant class of type fc. 

Determining the luminosity class of the secondary star is more difficult. We checked the spectra for signs of \niv\,\l\,4058\AA\ \citep{wal02}: this line appears as a very weak emission, much weaker than \niii\,\l\,4634,4641\AA, but the spectra are so noisy in the blue range that it is difficult to perform disentangling, hence to say whether the line belongs to one component or to both. In either case, this suggests that the hotter component (the secondary) is O3.5V or O4III-V and not a supergiant, whereas the presence in its spectrum of \niii\ and \heii\,\l\,4686\AA\ lines in emission suggests the contrary!

We used the typical stellar parameters of giant O3.5 and O5.5 stars \citep[ temperature and gravity are similar for giants and supergiants in this reference]{mar05} as input for the atmosphere modeling code CMFGEN. This enabled us to determine the equivalent widths of the \hei\,\l\,4471\AA\ and \heii\,\l\,4542, 5412\AA\ lines on the synthetic spectra, which we compared to the values found by fitting Gaussians to the observed spectra (at maximum velocity separation) or to the disentangled spectra. This led to nearly equal luminosity ratios, while the orbital solution further suggests equal masses. Therefore, our favored spectral types are O5--5.5I for the primary and O3--4III for the secondary.

Our monitoring of \cyg\ allows us to search for spectral signatures of a wind-wind collision. To this aim, one generally turns to Balmer and \heii\,\l\,4686\AA\ emissions. As already mentioned, the \heii\,\l\,4686\AA\ line has a peculiar behavior: its shape is relatively constant, but its RVs are shifted relative to the stars (Fig. \ref{rvcurve}).  The recorded velocities instead suggest the line arises close to the secondary star (but not close to its photosphere, because of the velocity shift), and a contamination by the collision zone is possible but difficult to assess. 

In contrast, the broad and intense \ha\ emission line displays no obvious shift with orbital phase but its profile clearly varies though the changes are not dramatic (Fig. \ref{ha}).  We first describe these variations. In this context, one has to keep in mind that only a single periastron passage could be observed in details, that of 2011, so that caution must be applied when generalizing the observations with $\phi\sim0$ to the usual behavior of the star. At phases $\phi > 0.1$, the line displays an overall triangular shape with a basis covering about 2000\,\kms\ and a peak at $\sim$200\,\kms. Superimposed on this emission are two relatively narrow absorptions with velocities of about $-175$ and 45\,\kms\ (i.e., not stellar velocities). It should be noted that the RVs of these two absorptions do not seem to vary much, but neither do the stellar RVs at these phases. As time goes by, the emission peaks decrease slightly in strength, though, as do the absorptions (the blueshifted absorption disappearing only after the redshifted one). Before periastron ($\phi=0.85-0.98$), the line thus appears as a nearly perfect triangle. Suddenly, shortly before periastron (at about $\phi=0.99$ to the best of our knowledge - our data do not enable us to further pinpoint the exact time of the event), the emission peak becomes more intense and is shifted to the red. It now appears at about 300\,\kms. In parallel, a strong absorption appears at $\sim -200$\,\kms, and is also much broader than before, since it is approximately 60\% wider than the previous blueshifted absorptions. A substructure of this absorption may be related to the photospheric absorption by the primary star, though this component is not obviously detected at other phases. In the WIRO data, the emission appears strongest at $\phi=0.999$, whereas the absorption is most intense at $\phi=0.998$, but this cannot be confirmed by Sophie data owing to the lack of coverage of these phases. After periastron, the emission peak decreases in strength and moves to less blueshifted velocities, while the absorption becomes less broad, and a second absorption gradually appears, coming back to the start of the cycle. 

A very broad \ha\ emission is quite compatible with a line arising not at the photosphere, but over a wide area - maybe the wind-wind collision in our case. If the wind-wind collision zone is dense enough at periastron, it could absorb some light from the stellar photospheres, in a kind of atmospheric eclipse. In such a case, the broad absorption appearing at periastron may be related to absorption of the primary's light by the colliding wind zone (Fig. \ref{sketch}). Indeed, the collision zone may well be highly distorted at periastron by Coriolis effects \citep[e.g. ][]{par08}, and one of its ``wings'' could then obscure our view of the primary, as seen from Earth. In such a configuration, the emission would come from the other zones of the collision, which are mostly redshifted and do not appear in front of one star, as seen from the observer's point of view. The pair of narrower absorptions seen at other phases is more difficult to explain, since the density of the collision zone decreases strongly with increasing separation, but they could be explained by a similar effect if there is enough dense cold gas downstream. Their constant velocity, though, is not surprising since the stars themselves do not show much RV change at these phases. The disappearance of all structures at $\phi= 0.85-0.98$ is, however, puzzling. It could be related to a physical change in the collision zone, or be simply due to viewing angle effects.

\section{\cyg, as seen in X-rays}
\subsection{The spectra}
\begin{figure*}[htb]
\includegraphics[width=6.2cm, bb=25 150 550 700, clip]{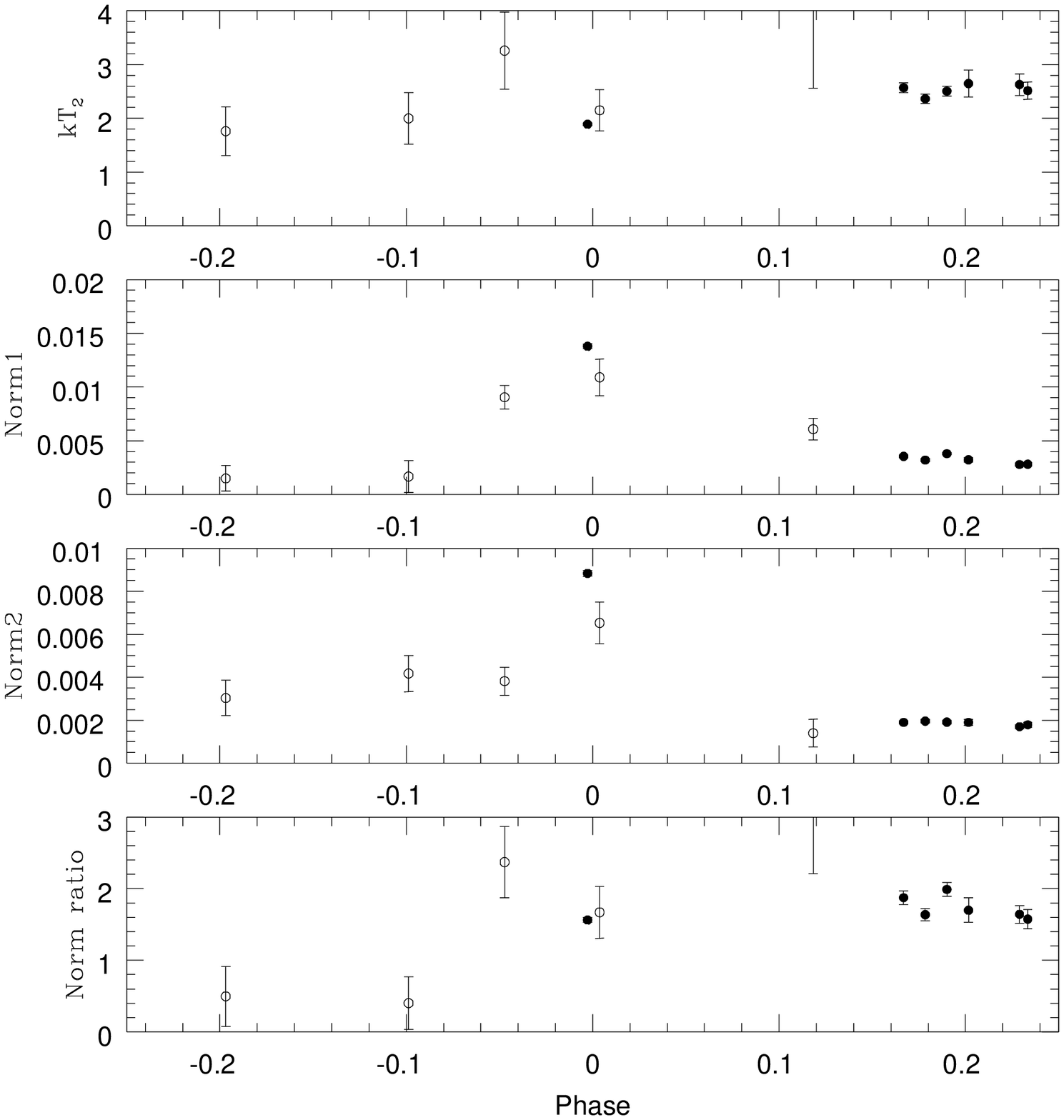}
\includegraphics[width=6.2cm, bb=18 145 550 640, clip]{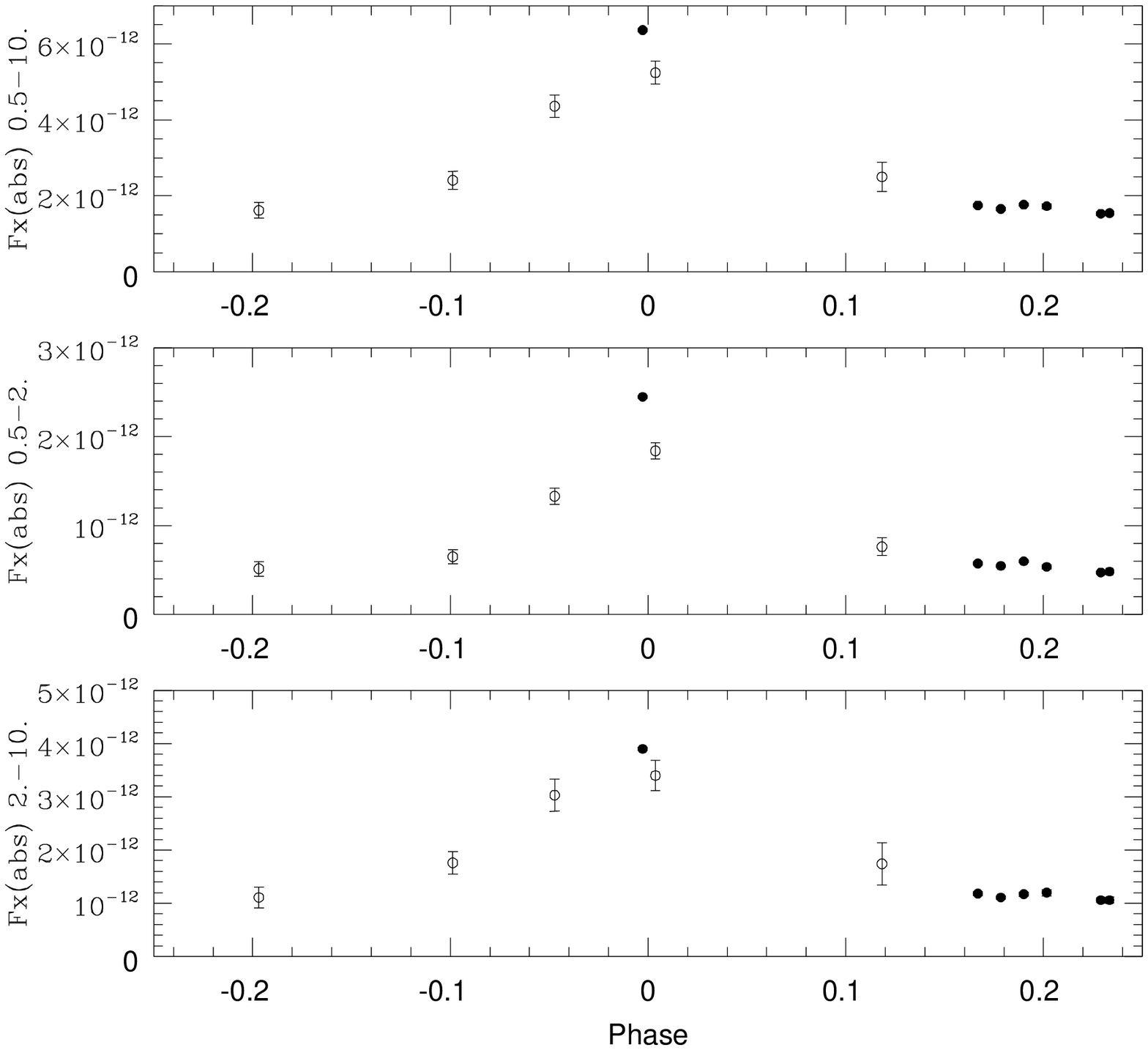}
\includegraphics[width=6.2cm, bb=50 170 550 675, clip]{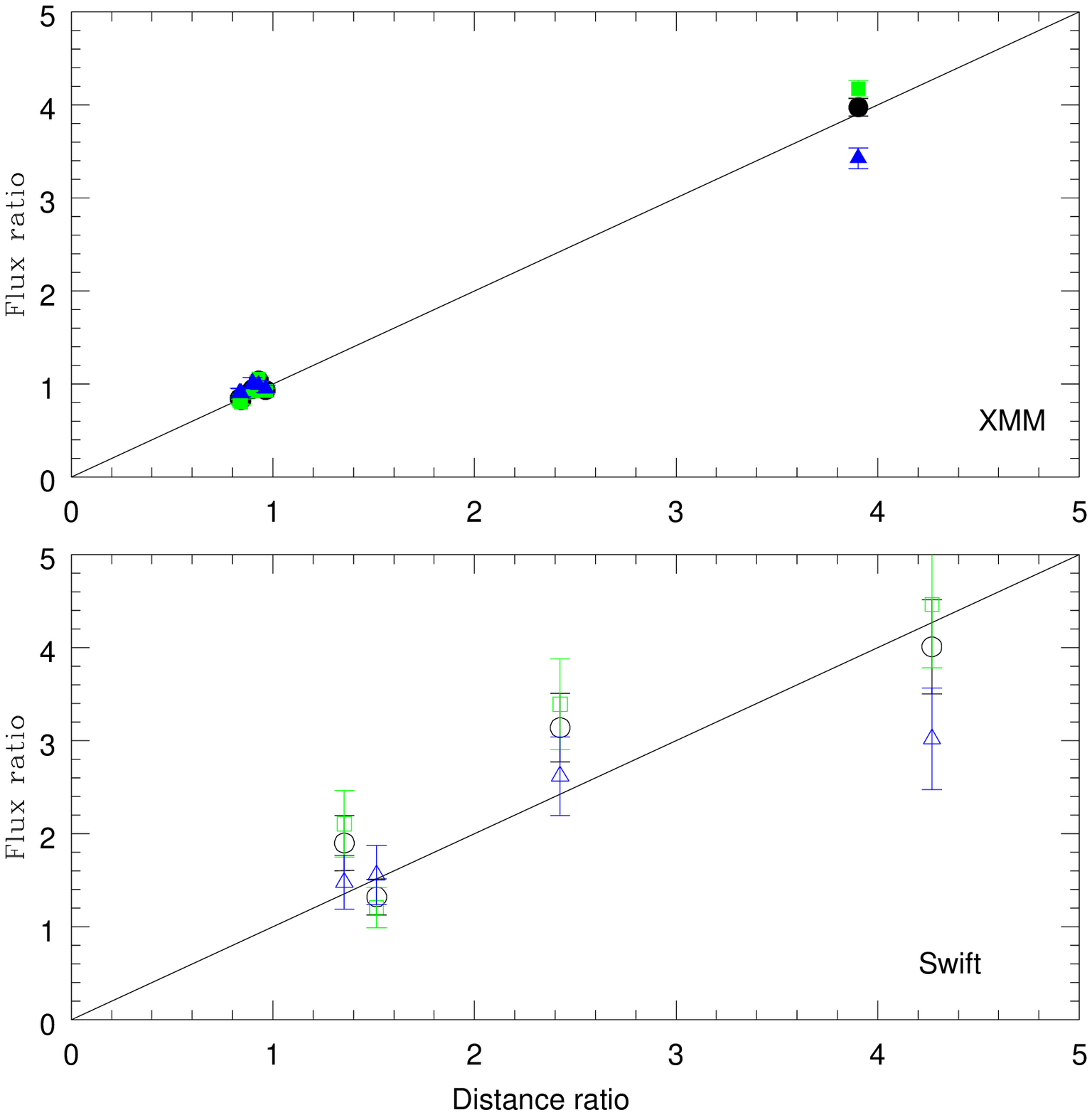}
\caption{Variations in the spectral parameters (left) and observed flux (middle) with phase, for the best-fit model with the first temperature and the absorptions fixed. The solid and open symbols correspond to \xmm\ and {\it Swift} data, respectively. The right panel shows, for the same model, the flux ratio $F^{unabs}_{X}/F^{unabs}_{X, ref}$ as a function of the inverse stellar separation ratio $D_{ref}/D$ for the \xmm\ and {\it Swift} data. The reference data corresponds to the first observation of each dataset, i.e., data from Oct\,04 for \xmm\ and data from Jan\,11 for {\it Swift}. The black circles corresponds to the total, 0.5--10.0\,keV flux, the green squares to the soft, 0.5--2.0\,keV flux, and the blue triangles to the hard, 2.0--10.0\,keV flux.}
\label{xray}
\end{figure*}

First, we fitted the new \xmm\ spectra and the old ones with optically-thin thermal plasma models of the form $wabs*phabs*(apec+apec)$. The fitting was performed within Xspec v12.4.0ad (which uses APEC 1.3.1): the first absorption was fixed to the interstellar one, $1.15\times10^{22}$\,cm$^{-2}$, and the second one allows for additional, in-situ absorption by the winds\footnote{It may be noted that due to the high interstellar extinction, there is no usable data at the lowest energies: using an ionized or a neutral absorption for the wind therefore does not change the results, since they mostly differ below 1\,keV.}; two thermal emission components were necessary to obtain a good fit. Because these \xmm\ data display the higher signal-to-noise ratios, this fitting enables us to see whether temperatures and absorption change with phase. The values found for the additional absorption and the cooler temperature in the different datasets were found to remain within one sigma of each other, so that we decided to fix them. {\it Swift} data have fewer counts, hence are noisier. A fully free 2T fitting results  in large error bars and sometimes erratic variations. It was thus not possible to check the absorption and cooler temperature constancy on these data. Therefore, all {\it Swift} spectra were fitted with the cooler temperature and additional absorption fixed. This seems a reasonable assumption since (1) the cooler component has a temperature typical of the intrinsic emission of hot, massive stars \citep[see e.g.][and references therein]{naz09}, and (2) the additional absorption is probably associated to the stellar winds, and their observed densities should not change much with phase since the two stellar components are similar (unlike what happens in WR+O systems). The contribution from a wind-wind collision should occur over a wider temperature range, and thus be the sole one responsible for the hotter component, and it should also be more variable than intrinsic emissions, especially in highly eccentric systems (see below). Table \ref{xfit} yields the results of these fits, and Fig. \ref{xray} illustrates the variations of the parameters with orbital phase.

We also tried to fix the  hotter temperature further, but the $\chi^2$ was significantly larger for the \xmm\ data (Table \ref{xfit}). However, the flux changes were similar for both \xmm\ and {\it Swift}, so that the conclusions reported in the next section are not affected by this choice. 

Two datasets were taken close to periastron, 5.5 days apart, one by \xmm\ and one by {\it Swift}. They show a difference in flux of about 20\%. The cross-calibration between {\it Swift} and \xmm\ data is believed to be better than 20\% (A. Beardmore, private communication, see also \citealt{tsu11}). To check our results, we looked at the other sources in the {\it Swift} field-of-view. Cyg\,OB2\,\#5 and 12 are known to vary, but unfortunately, their variations are not fully explained and predictable \citep{lin09,rau11}. The behavior of the colliding-wind binary Cyg\,OB2\,\#8A is, however, well known \citep{deb06,blo10}, and the {\it Swift} data are here fully consistent with the \xmm\ and {\it ASCA} data, when taking the errors due to a lower number of counts in the {\it Swift} data into account. The rapid brightening of \cyg\ near periastron, on timescales of days, may thus well be real.

\subsection{Archival data}
The Cyg\,OB2 association has been observed by all major X-ray facilities since its main members were detected in X-rays in 1978 by {\it Einstein}. To see whether our conclusions on \cyg\ are valid, we also considered these old datasets. 

{\it Einstein} IPC count rates of Cyg\,OB2 have been reported by \citet[see also the {\it Einstein} IPC catalog available at HEASARC]{wal98}: they amount to $0.0257\pm0.0018$\,cts\,s$^{-1}$ in mid-December 1978 ($\phi=0.16$), $0.0237\pm0.0042$\,cts\,s$^{-1}$ on 1979 Nov 22 ($\phi=0.56$), and $0.0322\pm0.0014$\,cts\,s$^{-1}$ on 1980 Jun 9 ($\phi=0.79$). However, \citet[ HEASARC ``ipcostars'' catalog]{chl89} quote an IPC count rate of $0.0406\pm0.0010$\,cts\,s$^{-1}$, which is not compatible with the previous values though they come from the same data. It is thus difficult to know exactly what happened 30 years ago. 

{\it ROSAT} observed Cyg\,OB2 in April 1991 and 1993 with the PSPC (ObsID=200109 and 900314, respectively) and in November 1994 and April 1995 with the HRI (ObsID\,=\,201845 parts 1 and 2, respectively). Spectra of the source (reduced to a single bin for HRI) were extracted using the same regions as for {\it Swift} observations. In these regions, the {\it ROSAT} background-corrected count rates of \cyg\ amount to $(3.96\pm0.36)\times10^{-2}$\,cts\,s$^{-1}$ in PSPC/1991, to $(3.83\pm0.15)\times10^{-2}$\,cts\,s$^{-1}$ in PSPC/1993, to $(2.3\pm0.2)\times10^{-2}$\,cts\,s$^{-1}$ in HRI/1994, and to $(1.47\pm0.06)\times10^{-2}$\,cts\,s$^{-1}$ in HRI/1995. The PSPC values are compatible with the PSPC count rates reported by \citet{wal98}\footnote{However, \citet{wal98} report an observed flux in the 0.1--3.5\,keV range of $\sim6\times10^{-13}$\,erg\,cm$^{-2}$\,s$^{-1}$. We instead found a flux of $\sim9\times10^{-13}$\,erg\,cm$^{-2}$\,s$^{-1}$ with our best-fit model. The discrepancy may come from a typo, as it is well known that {\it ROSAT} was not sensitive beyond 2.5\,keV: if we take 2.0 or 2.5\,keV as the upper limit of the energy range, rather than 3.5\,keV, Waldron's value and ours indeed agree.}. We furthermore fitted the PSPC spectra using the same model as above, with temperatures and absorption fixed (the poor signal-to-noise and low sensitivity at high energies did not permit us to free more parameters). Fitting results are reported in Table \ref{xfit}. 

\clearpage 
\begin{sidewaystable}
\caption{Results of the X-ray spectral fitting.}
\label{xfit}
\centering
\begin{tabular}{lccc| cccccc | ccccc}
            \hline\hline
Date & HJD & $\phi$ & $D/a$ & \multicolumn{6}{c|}{Absorptions and one temperature fixed} & \multicolumn{5}{c}{ Absorptions and temperatures fixed}\\
& -2400000& & & $kT_2$ & $norm_1$ & $norm_2$ & $F_X^{obs}$ & $F_X^{unabs}$ & $\chi^2$ (dof) & $norm_1$ & $norm_2$ & $F_X^{obs}$ & $F_X^{unabs}$ & $\chi^2$ (dof)\\
 & & & & keV & 10$^{-3}$\,cm$^{-5}$ & 10$^{-3}$\,cm$^{-5}$ &\multicolumn{2}{c}{10$^{-12}$\,erg\,cm$^{-2}$\,s$^{-1}$} & & 10$^{-3}$\,cm$^{-5}$ & 10$^{-3}$\,cm$^{-5}$ &\multicolumn{2}{c}{10$^{-12}$\,erg\,cm$^{-2}$\,s$^{-1}$}& \\
\hline
\multicolumn{4}{l|}{\xmm} &&&&&&&&&&& \\
Oct 04  & 53308.580  &  0.167 & 1.13 & $2.57\pm0.09$ & $3.56\pm0.12$ & $1.90\pm0.07$ & $1.75\pm0.04$ & $5.31\pm0.11$ & 1.16 (394) & $2.80\pm0.10$ & $2.42\pm0.06$ & $1.60\pm0.03$ & $4.85\pm0.10$ & 1.37 (395) \\
Nov 04a & 53318.558  &  0.178 & 1.17 & $2.37\pm0.09$ & $3.21\pm0.12$ & $1.96\pm0.08$ & $1.66\pm0.03$ & $4.96\pm0.10$ & 1.19 (422) & $2.60\pm0.10$ & $2.38\pm0.06$ & $1.55\pm0.03$ & $4.62\pm0.09$ & 1.30 (423) \\
Nov 04b & 53328.544  &  0.190 & 1.21 & $2.51\pm0.09$ & $3.80\pm0.11$ & $1.91\pm0.07$ & $1.77\pm0.04$ & $5.52\pm0.11$ & 1.28 (417) & $3.13\pm0.10$ & $2.39\pm0.06$ & $1.63\pm0.03$ & $5.13\pm0.10$ & 1.44 (418) \\
Nov 04c & 53338.506  &  0.202 & 1.25 & $2.65\pm0.25$ & $3.23\pm0.22$ & $1.90\pm0.14$ & $1.73\pm0.06$ & $5.03\pm0.18$ & 1.15 (125) & $2.42\pm0.18$ & $2.48\pm0.10$ & $1.57\pm0.06$ & $4.54\pm0.16$ & 1.39 (126) \\
Apr 07  & 54220.355  &  0.229 & 1.34 & $2.63\pm0.20$ & $2.79\pm0.14$ & $1.70\pm0.10$ & $1.53\pm0.05$ & $4.40\pm0.14$ & 1.08 (214) & $2.17\pm0.12$ & $2.15\pm0.07$ & $1.38\pm0.04$ & $4.00\pm0.12$ & 1.29 (215) \\
May 07  & 54224.170  &  0.233 & 1.35 & $2.52\pm0.16$ & $2.82\pm0.17$ & $1.79\pm0.11$ & $1.55\pm0.05$ & $4.48\pm0.15$ & 0.84 (158) & $2.23\pm0.14$ & $2.19\pm0.08$ & $1.41\pm0.05$ & $4.10\pm0.14$ & 0.98 (159) \\
Jun 11  & 55738.255  &  0.997 & 0.29 & $1.89\pm0.02$ & $13.8\pm0.22$ & $8.83\pm0.13$ & $6.36\pm0.03$ & $21.1\pm0.1$  & 1.59 (444) & $13.8\pm0.16$ & $8.83\pm0.08$ & $6.36\pm0.03$ & $21.1\pm0.1$  & 1.59 (445) \\
\hline
\multicolumn{4}{l|}{{\it Swift}}&&&&&&&&&&& \\
Jan 11 & 55571.618  & 0.803 & 1.24 & $1.76\pm0.45$ & $1.51\pm1.20$ & $3.04\pm0.82$ & $1.62\pm0.21$ & $4.14\pm0.54$ & 0.53 (18) & $1.93\pm0.73$ & $2.81\pm0.40$ & $1.66\pm0.21$ & $4.32\pm0.56$  & 0.51 (19) \\
Apr 11 & 55655.837  & 0.901 & 0.82 & $2.00\pm0.48$ & $1.68\pm1.50$ & $4.17\pm0.84$ & $2.41\pm0.24$ & $5.47\pm0.54$ & 1.45 (25) & $1.47\pm0.85$ & $4.38\pm0.50$ & $2.36\pm0.23$ & $5.18\pm0.51$  & 1.40 (26) \\
May 11 & 55700.082  & 0.953 & 0.51 & $3.26\pm0.72$ & $9.05\pm1.10$ & $3.82\pm0.65$ & $4.36\pm0.30$ & $13.0\pm0.9$  & 1.06 (47) & $6.25\pm1.10$ & $5.69\pm0.58$ & $3.71\pm0.25$ & $11.1\pm0.8$   & 1.29 (48) \\
Jul 11 & 55743.839  & 0.004 & 0.29 & $2.15\pm0.38$ & $10.9\pm1.7$  & $6.53\pm0.98$ & $5.24\pm0.30$ & $16.6\pm0.9$  & 0.93 (54) & $9.53\pm1.30$ & $7.48\pm0.68$ & $5.07\pm0.29$ & $15.8\pm0.9$   & 0.93 (55) \\
Oct 11 & 55842.169  & 0.118 & 0.91 & $6.16\pm3.60$ & $6.09\pm1.00$ & $1.40\pm0.65$ & $2.50\pm0.38$ & $7.87\pm1.20$ & 0.55 (17) & $4.31\pm0.99$ & $2.59\pm0.50$ & $1.89\pm0.29$ & $6.44\pm0.98$  & 0.78 (18) \\
\hline
\multicolumn{4}{l|}{{\it ROSAT}} &&&&&&&&&&&\\
Apr 91 & 48368.064  & 0.41 & 1.67 & & & & & & & $2.48\pm1.10$ & $2.03\pm1.20$ & $0.49\pm0.05$ & $3.13\pm0.33$ & 0.65 (11) \\
Apr 93 & 49109.312  & 0.28 & 1.46 & & & & & & & $2.93\pm0.44$ & $1.30\pm0.44$ & $0.45\pm0.02$ & $3.21\pm0.14$ & 0.97 (58) \\
\hline
\multicolumn{4}{l|}{{\it ASCA}} &&&&&&&&&&&\\
Apr 93 & 49106.958  & 0.27 & 1.45 & & & & & & & $1.30\pm0.31$ & $1.71\pm0.18$ & $1.03\pm0.06$ & $2.79\pm0.17$ & 0.96 (199) \\
\hline
\end{tabular}
\tablefoot{Fitted models are of the form {\it wabs$\times$phabs$\times$(apec+apec)}. The absorptions are fixed to 1.15 and $0.29\times 10^{22}$\,cm$^{-2}$. In the first case, the first temperature is fixed to 0.617\,keV, whereas in the second case, both temperatures are fixed ( cooler component at 0.617 and hotter component at 1.89\,keV). Fluxes refer to the  0.5--10.0\,keV band, except for {\it ROSAT} data where they are given in the 0.5--2.0\,keV band. }
\end{sidewaystable}
\clearpage

Both PSPC datasets were taken far from periastron, at phases 0.41 (in 1991) and 0.28 (in 1993), so that we expected few changes in recorded intensity. Indeed, the count rates, fluxes, and spectral parameters of the two observations agree with each other within the errors. The 1993 data was also taken at a similar phase as the sixth \xmm\ observation, and there is also fair agreement between them. HRI observations occurred just before and just after a periastron passage at phases 0.95 for the 1994 observation (i.e., similar in phase to the third {\it Swift} observation) and 0.11 for the 1995 observation (i.e., similar in phase to the fifth {\it Swift} observation). In fact, the recorded count rates are higher in the 1994 data, which is closest to the periastron time. 

Using the HRI response matrix and the same model as in the last \xmm\ observation (i.e., same temperatures, absorptions, and normalization ratio), we found an observed flux in the 0.5--2.\,keV band of 8 and $5\times10^{-13}$\,erg\,cm$^{-2}$\,s$^{-1}$ for the 1994 and 1995 observations, respectively. These fluxes agree well with the ones observed at similar phases with {\it Swift}. Overall, the {\it ROSAT} data thus appear compatible with the variations seen in \xmm\ and {\it Swift} observations.

{\it ASCA} observed Cyg\,OB2 in April 1993, at phase 0.27, nearly at the same time as the second {\it ROSAT}-PSPC observation. We extracted SIS spectra for both SIS-0 and SIS-1, and for both bright and bright2 modes (i.e., 4 spectra in total, implying that we merged data taken with different bit rates). For the source, we used the same region as in {\it Swift} and {\it ROSAT}, whereas the background region was chosen as a nearby circle of the same radius and on the same CCD. The source region is quite small for the {\it ASCA} resolution, but \cyg\ appears close to a CCD edge. Another problem in this observation is the high background: first, straylight from the nearby extremely bright source Cyg X-3 is contaminating the field-of-view; second, \cyg\ appears on the PSF wings of the bright Cyg\,OB2\,\#8A. Indeed, \cyg\ is not detectable by eye in single datasets, and it appears only as a faint object in the SIS global image. The resulting spectra are thus noisy, but we nevertheless tried to fit them with the same model as above (with temperatures and absorption fixed, see Table \ref{xfit}). The observed flux, in the 0.8-4.\,keV range amounts to $(8.4\pm0.5)\times10^{-13}$\,erg\,cm$^{-2}$\,s$^{-1}$, slightly smaller but within the errors of the value of $(1.1\pm0.4)\times10^{-12}$\,erg\,cm$^{-2}$\,s$^{-1}$ reported by \citet{kit96}. Compared to the sixth \xmm\ dataset (obtained at a similar phase) or the contemporary {\it ROSAT} observation, we again find a smaller flux in our {\it ASCA} fitting. However, considering the background and resolution problems, the agreement is not that bad.

In summary, the archival data agree well with the \xmm\ and {\it Swift} results, providing further support to the presence of phase-locked variations in \cyg\ since these different datasets were taken several periods apart.

\subsection{A wind-wind collision in \cyg}

The presence of bright and hard X-ray emission (with $kT>1.5$\,keV) is quite unusual in O-type stars. It may be linked to a magnetically-confined wind in a single star or to a wind-wind collision in a binary system. In our case, even though the nonthermal radio emission implies the presence of a magnetic field, the latter explanation seems to be the correct one. Indeed, \cyg\ displays long-term variations,  apparently compatible with the timescale of the orbital period.

Theoretical considerations show that there are two main sources of phase-locked variability amongst colliding wind binaries \citep[ see also \citealt{gue09} for a review of recent observations]{ste92}. The first one is the absorption. Since the collision is seen alternatively through the wind of each component, changes in absorption are expected. This is especially relevant for systems with two very different winds, e.g. WR+O: two spectacular cases in this respect are $\gamma^2$\,Vel \citep{wil95} and WR22 \citep{gos09}. The similarity of the two components in \cyg\ makes such variations less prominent; indeed, no significant change in absorption is detected in the data. 

\begin{table}
\caption{Distance and flux ratios for \xmm\ and {\it Swift} data, for the fits with the second temperature free to vary and with the observations taken in Oct04 and in Jan11 as references, respectively.}
\label{oneoverd}
\centering
\begin{tabular}{lcccc}
            \hline\hline
Date & $D_{ref}/D$ & \multicolumn{3}{c}{$F_X/F_{X, ref}$}\\
& & 0.5--10.\,keV & 0.5--2.\,keV & 2.--10.\,keV \\
            \hline
\multicolumn{4}{l}{\xmm} \\	 
Nov 04a & 0.96 & $0.93\pm0.03$ & $0.93\pm0.03$ & $0.95\pm0.04$ \\
Nov 04b & 0.93 & $1.04\pm0.03$ & $1.06\pm0.03$ & $0.99\pm0.04$ \\
Nov 04c & 0.90 & $0.95\pm0.04$ & $0.92\pm0.04$ & $1.01\pm0.06$ \\
Apr 07  & 0.84 & $0.83\pm0.03$ & $0.80\pm0.03$ & $0.91\pm0.04$ \\
May 07  & 0.83 & $0.84\pm0.03$ & $0.82\pm0.03$ & $0.91\pm0.05$ \\
Jun 11  & 3.90 & $3.97\pm0.09$ & $4.17\pm0.09$ & $3.43\pm0.11$ \\
\hline	 
\multicolumn{4}{l}{{\it Swift}} \\
Apr 11 & 1.51 & $1.32\pm0.19$ & $1.21\pm0.22$ & $1.56\pm0.32$ \\
May 11 & 2.43 & $3.14\pm0.37$ & $3.39\pm0.49$ & $2.62\pm0.42$ \\
Jul 11 & 4.27 & $4.01\pm0.51$ & $4.46\pm0.68$ & $3.02\pm0.55$ \\
Oct 11 & 1.35 & $1.90\pm0.30$ & $2.11\pm0.36$ & $1.48\pm0.29$ \\
\hline
\end{tabular}
\end{table}

The second source of phase-locked variability is the changing separation in eccentric binaries \citep{ste92}. For adiabatic cases, the separation between the stars is high, and the X-ray luminosity is expected to follow a $1/D$ relation (i.e., it should be maximum at periastron) with a minimal impact from the velocity changes because the winds are close to their maximal speeds at all phases. For radiative cases, the X-ray luminosity and plasma temperatures should be minimum at periastron, as the winds then collide at lower speeds. Adiabatic systems should be those with long periods, like \cyg, but not many long-period binaries have been monitored in X-rays, and results were mixed. $\gamma^2$\,Vel and WR22 did not show any $1/D$ variation, despite the changing separation of their components \citep{rau00b,gos09}. Cyg\,OB2\,\#8A and WR140 deviate from expectations at periastron, since the collisions then become radiative \citep{deb06,pol02,deb12}. Hints of the influence of a $1/D$ relation may have been found in HD\,93205 \citep[see Fig. 2 in][]{ant03}, though the results need confirmation, in view of the huge noise, and in HD\,93403, though the intrinsic emission of the stars blurred the results, yielding a luminosity change lower than that expected from a simple $1/D$ relation \citep{rau02}. Among evolved systems, WR25 seems to display a rather ``clean'' $1/D$ variation in its luminosity \citep{gos07,gue09}. For most of its orbit, it is also the case of $\eta$\,Car \citep{oka08}, though the system shows large cycle-to-cycle variations in flux, and there are strong indications that the wind collision region collapse onto the companion star at periastron \citep[and references therein]{par11c}. Up to now, no comparable example among O+O binaries has been found, so \cyg\ may well become the first one.

Indeed, while there seems to be no change in absorption, at least for our sampling, the normalization factors and fluxes clearly change with time. They show a sharp increase near periastron (see middle panel of Fig. \ref{xray}). In fact, a strong correlation appears when the fluxes are compared to the inverse of the stellar separation. The correlation is not perfect, but we know that there are instrumental differences (up to 20\% in flux, see above) and larger errors for the less sensitive {\it Swift} data. To reduce the impact of these instrumental effects, we performed a comparison within each dataset (see last panel of Fig. \ref{xray}  and Table \ref{oneoverd}): within the errors, the fluxes vary as $1/D$, as expected from an adiabatic collision in an eccentric system \citep{ste92}.  In fact, the linear correlation coefficient between the distance and flux ratios listed in Table \ref{oneoverd} is 0.97 for Swift+\xmm\ data and 0.999 for the more precise \xmm\ data alone. In this context, it may also be noted that the increase by a factor of 1.6 in the {\it ROSAT}-HRI count rates is well matched by the decrease in separation between orbital phase $\phi=0.11$ and $\phi=0.95$ by the same factor. The only slight departure from the $1/D$ relation (Fig. \ref{xray}) is the hard X-ray flux appearing slightly below the expected value: this may be due to the collision becoming slightly radiative because of the increased density at periastron. This would also explain the presence of cool plasma responsible of \ha\ absorption at these phases in the optical spectra. However, this does not affect the bulk of the X-ray emission, since the $1/D$ correlation is closely followed overall, with no evidence of blurring by an additional, constant intrinsic emission from the stars. The absence of dramatic changes in \ha\ also correlates well with an adiabatic nature for the wind-wind collision.

On the other hand, the second temperature of the spectral fitting is clearly lower near periastron in the \xmm\ data (see left panel of Fig. \ref{xray}). This may be explained by a shock occurring at a smaller separation. Indeed, a shorter distance between the two stars may imply that winds collide before they could accelerate to full speed, resulting in a cooler shocked plasma. Moreover, at small separations, radiative braking/inhibition \citep{ste94,gay97} may take place, slowing down the winds further before they collide. Such a cooling of the hard component is predicted by models of wind-wind collisions in eccentric binaries: \citet[see their Fig. 22]{pit10} find a change by a factor of 3 in the highest temperature, whereas \citet{par11} find an order of magnitude variation when the shock plunges deep into the wind acceleration region. In \cyg, the temperature only changes by 25\%, but the separation is much larger than in the case modeled by \citet{pit10} and the wind-wind collision is closer to balance than in the case modeled by \citet{par11}. Therefore, we expect the influence of wind acceleration in \cyg\ to be less pronounced than found by these authors, weakening the cooling effects. 

\begin{figure}
\includegraphics[width=9cm]{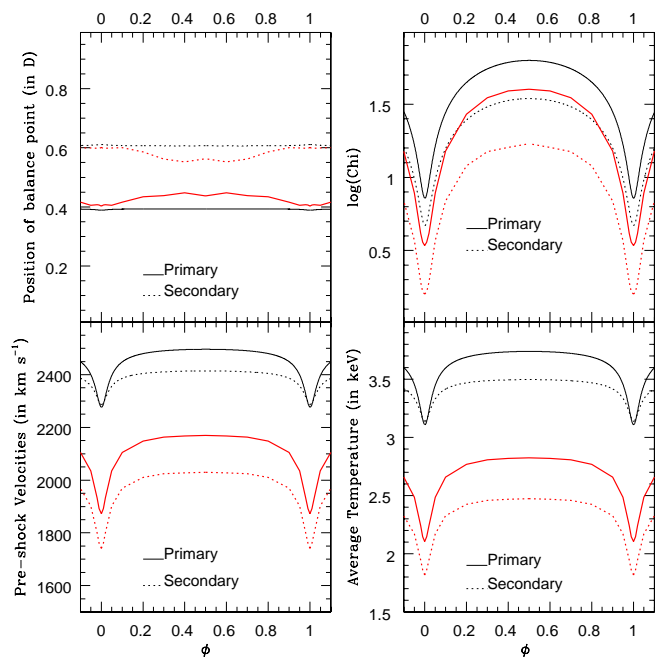}
\caption{Evolution with phase of the collision parameters (position of balance point, cooling parameter $\chi$, pre-shock velocities, and average plasma temperature, $kT=0.6\times v({\rm in\,\,1000\,km\,s^{-1}})^2$). Thick red lines and thin black lines correspond to the cases with and without radiative inhibition, respectively.}
\label{model}
\end{figure}

To put these observing results in perspective, we did a first modeling of the wind-wind collision in \cyg.  We here use the analytical models of \citet[see also a detailed example of the calculation in \citealt{par11b}]{ste92}. From the ram pressure balance, they allow us to find the position of the collision, the cone's opening angle, plasma temperatures (from the pre-shock wind velocities\footnote{We assume that the emission-weighted post-shock temperature is $kT({\rm in\,\,keV})=0.6\times v({\rm in\,\,1000\,km\,s^{-1}})^2$. This is roughly half the temperature at the stagnation point and is meant to account for the lower temperatures obtained in the downstream flow as the off-axis shocks become oblique.}), and the cooling parameter (to determine the nature of the collision, adiabatic if $>1$, radiative otherwise). As input, such models require wind and stellar parameters: stellar radii and separation, wind mass-loss rates, and velocities. The orbital solution has just been found, yielding the separation at each phase. However, not much information is available on the individual wind parameters, only approximate spectral types have been derived. We therefore made the following assumptions: inclination of the system of about 62 degrees \citep[so that the true masses are compatible with the predictions of ][]{mar05}, typical stellar parameters of O5-5.5I and O3-4III stars taken from \citet{mar05}, and mass-loss rates derived from the recipe of \citet{vin00}. This yields mass-loss rates of $5\times 10^{-6}$\,M$_{\odot}$\,yr$^{-1}$ and $1.2\times 10^{-5}$\,M$_{\odot}$\,yr$^{-1}$ and terminal velocities ($=2.6\times$escape velocities) of 2540\,\kms\ and 2440\kms\ for the primary and secondary, respectively. We also assume the usual $\beta$-law with $\beta=1$ for the wind velocity evolution. At maximum separation, this results in a wind momentum ratio of 0.43 (or 2.3 in favor of the secondary), which gives an opening angle of the collision zone of about 150 degrees, compatible with the radio map \citep[see Fig. 6 of ][ note that the angle estimated from that figure is approximate, because sensivity or wind opacity may affect its determination]{dou06}. Figure \ref{model} shows the evolution with phase of several parameters of the collision zone. The position of the balance point (at 0.4$D$ from the primary) does not change with phase, and the cooling parameter $\chi$ \citep{ste92} is always $>>1$, implying that the shocks should still be relatively adiabatic even at periastron (which nicely conforms with the X-ray observations). The agreement is not perfect, however, since the shocked plasma temperatures appear too high (about 3.6\,keV at apastron and 3.1\,keV at periastron vs observed values of 2.5\,keV and 1.9\,keV), and the relative change in temperatures is slightly too low (13\% vs 25\% observed). We therefore include the effect of radiative inhibition in the model: rather than using a simple $\beta$ velocity law, we here assume the winds to be radiatively driven following the \citet{cak} approximation and including the influence of the opposing star radiation field \citep[e.g. ][]{ste94,gay97}. Such a model improves the agreement with observations: the winds collide at lower speeds, even at apastron, resulting in lower plasma temperatures (about 2.7\,keV at apastron and 2.0\,keV at periastron), and the decrease in plasma temperature is now 25\% between apastron and periastron (Fig. \ref{model}). In this case, the shock still remains mostly adiabatic throughout the orbit, as observed. Wind clumping may play a further role in decreasing plasma temperatures \citep[and references therein]{pit07}. The next step will be to make full 3D hydrodynamical simulations of the wind-wind collision \citep[c.f.][]{pit09,par11}, to calculate synthetic spectra and lightcurves. Their comparison with observations will allow determination of the density, temperature, and velocity of the hot plasma within the wind-wind collision region, helping to constrain the stellar mass-loss rates and wind speeds in this system. Constraints on the magnetic field within the collision region will also be derived. However, this detailed modeling requires all data to be at hand (not only optical and X-ray observations - see future papers of the series); it is also clearly beyond the scope of this paper.

\section{Summary and conclusions}

In mid-2011 the first observable periastron passage of \cyg\ occurred since the  binarity discovery. To observe this event under all aspects, we organized a multiwavelength campaign with a simultaneous coverage at radio, infrared, optical, and X-ray energies. Here we report on the results of the optical and X-ray monitorings.

The optical spectroscopic monitoring obtained with the Sophie spectrograph enabled us to refine the orbital solution, confirming preliminary results. A spectral disentangling was also performed, revealing the two stars to be of O3--4III and O5--5.5I types for the secondary and primary, respectively. While the \heii\,\l\,4686\AA\ line follows the motion of the hottest component (though with a shift), the \ha\ line appears extremely broad but unshifted. The line profile changes with phase, with both enhanced absorption and stronger emission at periastron.

X-ray observations by \xmm\ and {\it Swift} have been obtained as \cyg\ reached the 2011 periastron. The high-energy monitoring revealed a rather constant X-ray spectrum, except at periastron where the emission becomes slightly softer. The second thermal component then has a temperature reduced by about 25\% at periastron, which is reproduced well by collision models including radiative inhibition. The monitoring also unveiled the luminosity variations of the system, which follow the $1/D$ changes predicted for adiabatic collisions remarkably closely. The only (slight) departure from that relation is a very small decrease in the hard-band flux, which could indicate that the collision has become very slightly radiative at periastron. Overall, \cyg\ seems the best (and the first!) example of an adiabatic O+O colliding-wind system.

\begin{acknowledgements}
The authors thank Tabetha Boyajian, Micha\"el De Becker, and Gregor Rauw for useful comments. We also thank the {\it Swift} PI, Dr Neil Gehrels, and the \xmm\ project scientist, Norbert Schartel, for having made the X-ray monitoring possible. YN acknowledges useful discussion with Kim Page on {\it Swift} data reduction and calibration. The Li\`ege team acknowledges support from the European Community's Seventh Framework Program (FP7/ 2007-2013) under grant agreement number RG226604 (OPTICON), the Fonds National de la Recherche Scientifique (Belgium), the Communaut\'e Fran\c caise de Belgique, the PRODEX XMM and Integral contracts, and the `Action de Recherche Concert\'ee' (CFWB-Acad\'emie Wallonie Europe).  JMP would like to thank the Royal Society for funding a University Research Fellowship. We thank the OHP staff and colleagues who took the service mode observations. We thank Michael Alexander, Garrett Long, Mike Lundquist, Jessie Runnoe, Rachel Smullen, Carlos Vargas, and Earl Wood for their assistance in obtaining the data at WIRO.  We thank WIRO staff James Weger and Jerry Bucher for their work that enabled this science.  ADS and CDS were used for preparing this document. 
\end{acknowledgements}

\end{document}